\documentstyle[12pt,epsf]{article}
\newcommand{\nc}{\newcommand}
\nc{\beq}{\begin{equation}}
\nc{\eeq}{\end{equation}}
\nc{\beqa}{\begin{eqnarray}}
\nc{\eeqa}{\end{eqnarray}}
\nc{\lsim}{\mbox{\raisebox{-.6ex}{~$\stackrel{<}{\sim}$~}}}
\nc{\gsim}{\mbox{\raisebox{-.6ex}{~$\stackrel{>}{\sim}$~}}}

\begin{document}
\input epsf.tex

\title{\vskip2cm The Sphaleron Rate:  B\"{o}deker's Leading Log}

\author{Guy D.~Moore\footnote{e-mail: 
        guymoore@phys.washington.edu} \hspace{3cm}\\
{\small Department of Physics, University of Washington, Seattle
        WA 98195-1560 USA } }
\maketitle

\begin{abstract}
B\"{o}deker has recently shown that the high temperature
sphaleron rate, which measures baryon number violation in the hot
standard model, receives logarithmic corrections to its leading
parametric behavior; $\Gamma = \kappa' [\log(m_D / g^2 T) + O(1)] (g^2
T^2 / m_D^2) \alpha_W^5 T^4$.  After discussing the
physical origin of these corrections, I compute the leading log
coefficient numerically; $\kappa' = 10.8 \pm 0.7$.  
The log is fairly small relative to the $O(1)$ ``correction;'' so
nonlogarithmic contributions dominate at realistic values of the
coupling.  
\end{abstract}

\vspace{-5in}

\rightline{MCGILL/98-28}
\rightline{hep-ph/9810313}

\vspace{5in}

\section{Introduction}

It has been known for some time now that baryon number is {\em not} a
conserved quantity in the minimal standard model \cite{tHooft}.  It is
violated nonperturbatively because of the anomaly, the chiral couplings
of fermions to SU(2) weak, and the topologically nontrivial vacuum
structure of SU(2).  However, as is characteristic of a nonperturbative
process in a weakly coupled theory, the rate of violation is so tiny
that it is completely irrelevant phenomenologically.  Certainly, if
baryon number is also violated due to high dimension operators descended
from some GUT, baryon number violating decay rates due to the GUT mechanism
will greatly exceed the electroweak rate, of order $m_W^{1} \exp(-4 \pi
/ \alpha_W) < 10^{-170} {\rm GeV}$ even before accounting for additional
suppression from powers of small CKM matrix elements and a high power of
$(m_{\rm proton} / m_W)$.

However, as first realized by Kuzmin, Rubakov, and Shaposhnikov in 1985,
the efficiency of standard model baryon number violation is very much
higher at finite temperature \cite{KRS85}.  A perturbative estimate
\cite{ArnoldMcLerran} based on a saddlepoint expansion about Klinkhamer
and Manton's sphaleron \cite{Manton} indicated
that the rate is more than enough to erase any relic abundance of baryon
number left over from the GUT scale, unless the combination $B-L$,
baryon minus lepton number, is nonzero; this quantity is preserved by
electroweak physics.  It also may be that the baryon number abundance in
the universe was produced by electroweak physics, which motivates
the more careful study of electroweak baryon number violation at finite
temperature.   

Recently our understanding of thermal baryon number violation 
has improved, though it
is not complete.  We understand that the efficiency of baryon number
violation can in most relevant circumstances be related by a fluctuation
dissipation argument to the diffusion constant for Chern-Simons number
\cite{KhlebShap,Mottola,RubShap}.  
This diffusion constant, called the sphaleron rate, is defined as
\beq
\Gamma \equiv \lim_{V \rightarrow \infty} \lim_{t \rightarrow \infty} 
	\frac{ \langle ( N_{\rm CS}(t) - N_{\rm CS}(0) )^2 \rangle}
	{Vt} \, ,
\eeq
where the expectation values refer to a trace over the equilibrium
thermal density matrix.  The quantity $\Gamma$ is the topological
susceptibility of the electroweak sector at finite temperature, in
Minkowski time.  

We know that, when the electroweak phase
transition is first order, the value of $\Gamma$ jumps discontinuously
between the phases; it is quite small in the broken phase and much
larger in the symmetric phase.  
An old power counting argument says that
the symmetric phase rate should be of order $\Gamma \sim \alpha_W^4 T^4$
with an order unity coefficient.  This argument relied, correctly, on
the natural nonperturbative length scale in the hot plasma being $1 / 
(\alpha_W T)$.  One then assumes that the natural time scale is the
same; on dimensional grounds the spacetime rate of $N_{\rm CS}$
diffusion must then be of order $(\alpha_W T)^3 \alpha_W T$.  
However, Arnold, Son, and Yaffe have shown that at leading parametric
order, the natural time scale is {\em not} $1 /
( \alpha_W T)$, but $1 / \alpha_W^2 T$ \cite{ASY}, up to possible 
logarithmic corrections, which the authors did not consider.  More
recently, B\"{o}deker has demonstrated that logarithmic corrections to
their argument do occur \cite{Bodeklog}.

This says nothing about the numerical value of the sphaleron rate; it
could be parametrically $\alpha_W^5 T^4$ but numerically irrelevantly
small.  Ambj{\o}rn and Krasnitz presented numerical evidence that it was
large, by considering classical, thermal Yang-Mills theory on the
lattice \cite{AmbKras}.  Their definition of $N_{\rm CS}$ was not
topological and could therefore suffer from potentially severe lattice
artifacts, but Turok and I studied the same system with a topological
definition of $N_{\rm CS}$ and verified that $\Gamma$ is substantial
\cite{slavepaper}.  The sphaleron rate we found, expressed in physical
units, was lattice spacing dependent, which turns out to be a prediction
of the arguments of Arnold, Son, and Yaffe; the extra power of
$\alpha_W$ arises from the interaction between the infrared fields and
ultraviolet excitations, as I will discuss more below, and on the
lattice it becomes one power of the lattice spacing $a$.  

In fact the lattice spacing dependence of $\Gamma$ for pure classical
lattice Yang-Mills theory only fits $\Gamma
\propto a$ if there are substantial corrections to scaling, which has
led some to call into question whether the Arnold, Son, and Yaffe's
analysis is correct.  In this paper I will assume that it is, as seems
justified on theoretical grounds and numerical evidence from classical
Yang-Mills theory ``enhanced'' with added degrees of freedom which
reproduce the hard thermal loop effects \cite{particles}.  I will return
to the large corrections to scaling in the classical lattice theory in
subsection \ref{sub_in_g}.

Because the interactions between infrared and ultraviolet excitations
are important to setting the sphaleron rate, it is somewhat difficult to
actually extract $\Gamma$ at leading order in $\alpha_W$ 
for the continuum quantum theory, at the physical value of $\alpha_W$ or
even in the parametric small $\alpha_W$ limit.  Hu and M\"{u}ller
proposed a technique based on including the UV physics, lost to the
lattice regulation, by introducing new degrees of freedom which
influence the IR fields in the same way\cite{HuMuller}.  They
implemented and applied the technique jointly with me \cite{particles}.
As I will discuss later, this technique still suffers from some poorly
controlled systematics, which are related to the logarithmic corrections
discovered by B\"{o}deker; in fact, beyond the leading log the infrared
physics the technique will simulate is not rotationally invariant.

A good first step to answering the remaining questions about the
sphaleron rate is to determine it at leading logarithmic order in
$\alpha_W$.  B\"{o}deker has demonstrated that this can be done within
an effective theory which is completely UV well behaved; in fact it is
nothing but the Langevin equation for 3-D Yang-Mills theory
\cite{Bodeklog}.  The leading log behavior is probably not very useful
by itself,  for estimating $\Gamma$ at the physical value of $\alpha_W$.
Leading log expansions often miss large constant corrections; we know
for instance that the $O(g^2 T)$ contribution to the Debye mass has a
much larger constant contribution than the leading log might suggest
\cite{KLRS_mD,LainePhilipsen}.  However, it is 
still useful to know the leading log; for instance, its size is related
to the severity of the systematic problems with the method of Hu and
M\"{u}ller, and it might in principle be useful for extrapolating
lattice results which correspond to an inappropriate value of $\alpha_w$
back to the correct value.

The purpose of this paper is to determine the coefficient of the leading
log behavior of $\Gamma$; namely, to find $\kappa'$ defined through
\beq
\Gamma = \kappa' \left( \log \frac{m_D}{g^2 T} + O(1) \right) 
	\frac{g^2 T^2}{m_D^2} \alpha_W^5 T^4 
	+ ( {\rm higher \; order}) \, .
\label{def_kappa}
\eeq
(My logs are always natural logs.)
Neglecting the $O(1)$ means that $\log (1/g)$ is treated as much larger
than any order unity constant, a rather extreme interpretation of the
perturbative expansion.  Probably this expansion is completely
unjustified, but as I said there are still important things to be
learned from making it.  Until Section \ref{interpretation} I will not
worry about whether the expansion in $\log (1/g) \gg 1$ is
justified; the goal is simply to determine $\kappa'$.  I will also work
in Yang-Mills theory, which is appropriate at leading order only for
temperatures well above the equilibrium temperature.  I will mention how
to include the Higgs field in Section \ref{interpretation}.

A summary of the paper is as follows.  Section \ref{review}
will discuss in an intuitive, physical, but nontechnical level why 
the rate has the parametric form I show; 
where the extra $\alpha$ comes from and particularly why there is a
log.  The section provides two apparently different arguments, one in
terms of conductivities and scattering processes for hard particles and
one in terms of hard thermal loops (HTL's) and Wilson lines; the two are of
course equivalent.  Some salient details about the Wilson line
are in Section \ref{Wilsonline}, which is more technical.  This section
also shows how the numerical model of Hu, M\"{u}ller and Moore fails
beyond leading log order.  Section \ref{numerics} studies B\"{o}deker's
effective theory for extracting the leading log,
numerically.  Since the Langevin dynamics are simple but
numerically costly, the emphasis is on controlling systematic errors.
Many of the details appear either in previous papers or in the appendix.
For the reader's ease I present the
answer now:  $\kappa' = 10.8 \pm 0.7$.  The dominant error here is
statistical; systematic errors are completely under control.  In Section
\ref{interpretation} I discuss the meaning of the result, and try to
estimate what the sphaleron rate is for the realistic values of $m_D^2$
and $g^2$ by using this result to extrapolate the results of Hu,
M\"{u}ller, and Moore to the right value of the log.  My estimate is
$\Gamma \simeq 20 \alpha_W^5 T^4$ for $m_D^2 = (11/6) g^2 T^2$ and $g^2
= 0.4$; however there are uncontrolled systematics which may be as large 
as $30\%$.  I also discuss corrections to the approximation that $g \ll
1$; the largest of these will be of order $10 \%$.  The 
conclusion concludes.  There is also a technical appendix which 
discusses the match between lattice and continuum Langevin time
scales; the match between continuum and Langevin time scales is computed
at $O(a)$ and the match between lattice Langevin and heat bath time
scales is determined by a measurement.

\section{The physics behind $\alpha_W^5 \log(m_D / g^2 T)$}
\label{review}

\subsection{note on the classical approximation}

To a very good approximation the behavior of infrared fields in 
thermal Yang-Mills or Yang-Mills Higgs theory at weak coupling is that
of classical fields.  The ``old'' parametric estimate, $\Gamma \propto
\alpha_W^4 T^4$, relies on the fact that the only length scale available
in classical Yang-Mills theory is $1 / g^2 T$.  It is also assumed that
the only time scale is the same, in which case any essentially infrared
spacetime rate must go as $ g^8 T^4$.  

This argument relies on a decoupling between the infrared and
ultraviolet fields, since the ultraviolet fields (by which I
mean fields of wave number $k \sim T$) do not behave
classically.
It is known that this decoupling is very accurate for
thermodynamic variables, except that the $A_0$ component of the gauge
field receives a Debye mass.  In fact the validity of the decoupling is
equivalent to the quality of the dimensional reduction approximation
\cite{oldDR}, which has been discussed extensively \cite{FKRS1,KLRS}.

However, the decoupling does not extend to dynamics; the generalization
of the Debye mass to unequal times are the hard thermal loops, which
significantly affect the infrared dynamics.  The hard thermal loops are
precisely that set of diagrams which are linearly divergent within the
classical theory.  This linear divergence is cut off at the ultraviolet
scale $k \sim \pi T$ where the theory ceases to behave
classically; so the size of these effects depends essentially on the way
the IR classical theory is regulated.\footnote{I should mention
parenthetically that the classical theory also contains quadratic and
cubic divergences in the energy density, but these do not affect the
IR dynamics responsible for baryon number violation.}
Thus it is only correct to say that the IR fields behave classically if
we mean that they behave like the IR fields of a classical theory
regulated in some way which correctly reproduces the hard thermal
loops.  In nature that regulator is quantum mechanics, but we might be
able to find some other appropriate regulator in a numerical setting.

\subsection{Argument in terms of conductivity and scatterings}

Here I will give an argument for the $\alpha_W^5 \log(1/g) T^4$ law
based on Lenz's Law and the conductivity of the plasma.  The argument
has very recently been made quantitative \cite{ASY2}, but I will present
it at the qualitative, intuitive level.

To see how hard thermal loops influence the sphaleron rate, I first make
the point that the sphaleron rate is set by the evolution of very soft
infrared fields, where by very soft I mean fields with wave number 
$k \sim g^2 T$.  Parametrically shorter wave lengths do not contribute
appreciably because the probability for nonperturbative physics to occur
at such scales is exponentially suppressed.  According to standard
sphaleron type arguments, the contribution from the
scale $k \sim g^{2-\delta} T$ is suppressed by of order 
$\exp(- k/g^2 T) = \exp(- g^{-\delta})$.  Even the scale 
$k \sim g^2 T \log (1/g)$ gives a contribution suppressed by a power of
$g$.  

The second point is that it is physics in the transverse sector which
matters, and in particular, diffusion of $N_{\rm CS}$ requires the
evolution of magnetic fields.  To see this, first go back to the
definition of $N_{\rm CS}$,\footnote{Although I set the speed of light
$c=1$, I typically write expressions noncovariantly with a positive
space metric, which is convenient in the finite temperature context
because the thermal bath establishes a preferred frame.}
\beq
N_{\rm CS}(t_2) - N_{\rm CS}(t_1) = \int_{t_1}^{t_2} dt \int d^3 x 
	\frac{g^2}{8 \pi^2} E_i^a B_i^a(x) \, .
\eeq
Now the $B$ field is always transverse, meaning that $D \cdot B = 0$,
by the Bianchi identity; and so
only the transverse part of the $E$ field contributes.  The Bianchi identity
also states that $\vec{D} \times \vec{E} = - [ D_t , \vec{B} ]$.  Since
the relevant part of the electric field is transverse, it will in
general have nonzero covariant curl.  For $E$ to be nonzero and to
remain the same sign for long enough to give a nontrivial contribution
to $\int E_i^a B_i^a dt$, there must then be time evolution of infrared
magnetic fields.

At this point it is useful to recall how infrared magnetic fields
evolve in the abelian theory, on wave lengths longer than the Debye
screening length.  The answer is familiar plasma physics; the plasma is
very conducting, and a conducting medium resists changes in magnetic
fields by Lenz's Law.  In the limit of infinite conductivity the
magnetic fields are perfectly frozen; for finite conductivity the time
scale for their evolution scales with conductivity.  
A magnetic field of wave number $k$, with well more
than its mean thermal energy density but much less energy density than
is contained in the bulk plasma, decays according to 
\beq
[D_0 , A_i] = \frac{dA_i }{dt}= - \frac{1}{\sigma(k,\omega \ll k)} 
	( k^2 \delta_{ij} - k_i k_j ) A_j \, ,
\eeq
in the parametric limit that the decay time is well longer than $1/k$,
which is satisfied at all length scales parametrically longer than the
Debye screening length.  The characteristic decay time of a magnetic
field in the plasma is then $\tau = \sigma(k , \omega \ll k) / k^2$.
Of course I have only written the dissipative part of the magnetic
field evolution equation; 
there must also be a noise term which is uniquely specified by
the requirement that the thermodynamics of the IR magnetic fields are
correct.  

Note that the conductivity is wave number dependent.  It has a good
infrared limit which is achieved for length scales larger than the mean
scattering length $l_{\rm free}$ of a current carried by hard particles.
In the abelian theory a particle's charge is preserved when it undergoes
a scattering, so this length scale is the mean length for
large angle scattering, parametrically $ l_{\rm free} \sim 
1 / \alpha^2 T$ times
logarithmic corrections.  If every particle had the same free path and
a scattering perfectly randomized its momentum, the conductivity on
scales longer than $l_{\rm free}$ would be 
\beq
\sigma = \frac{ m_D^2 l_{\rm free}}{3} \, , \qquad 
	k \ll 1/l_{\rm free} \, .
\label{sigma_long}
\eeq 
When scattering processes are more complicated this formula
defines an effective value of $l_{\rm free}$.  
For length scales well  between the Debye screening length and
$l_{\rm free}$, where scatterings of the charge carriers can be
neglected, the conductivity is related to the Debye
length through 
\beq
\sigma = \frac{\pi m_D^2}{4 k} \, , \qquad
	m_D \gg k \gg 1 / l_{\rm free} \, .
\label{sigma_intermed}
\eeq
(In both expressions the Debye length $m_D$, which is $O(gT)$, is just
keeping track of the number density, charge, and $\langle 1/E \rangle$ 
of the particles.The derivations of each expression assume
ultrarelativistic dispersion relations for the charge carriers.) 
This expression follows from the form of the transverse self-energy and
the fact that the conductivity we are discussing in this case is just a
special case of the HTL self-energy, $\sigma(k , \omega \ll k) = 
{\rm Im} \, \Pi_T(k , \omega) / \omega$.  This is the connection between
this ``conductivity'' picture and the hard thermal loops.

The behavior of the electroweak gauge fields for scales parametrically
between the nonperturbative scale $k = g^2 T$ and the Debye scale 
$k = g T$ is the same as in the abelian theory at leading parametric
order.\footnote{Some years ago Ambj{\o}rn and Olesen argued that
nonabelian fields obey an anti-Lenz's Law {\protect{\cite{freeon}}},
apparently in contradiction to the argument presented here.  Their work
refers to the nonabelian interactions between $W$ and $Z$ fields in the
presence of strong (electromagnetism) magnetic fields at zero
temperature.  It may have some bearing on the mutual interactions of the
fields at the $g^2 T$ scale, but the response of the harder modes to the
very soft fields is at leading order the Lenz law type behavior seen in
the abelian theory.}  Since the mean free path of a hard
excitation to undergo {\em any} scattering is $\sim 1 / ( g^2 T \log
(1/g))$\cite{Pisarski}, 
the conductivity for $k = g^{2 - \delta}T$, with $0 < \delta <
1$, is $\sigma = \pi m_D^2 / 4 k \sim g^\delta T$, and the time
constant associated with the decay of a magnetic field is $\tau \sim
1/(g^{4 - 3 \delta}T)$.  Although the scale $k \sim g^2 T$ 
does not fit within the range of validity of this argument it cannot be
that the decay rate for a magnetic field with $k \sim g^2 T$ differs
from the $\delta \rightarrow 0$ limit
by any nonzero power of $g$.  Hence the
relevant time scale for the dynamics of nonperturbative infrared
magnetic fields in weakly coupled, hot Yang-Mills theory is $O(( g^4
T)^{-1})$, up to corrections at most logarithmic in $g$.  This is a
paraphrase of the argument of Arnold, Son, and Yaffe \cite{ASY}, who
however neglected the possibility of logarithmic corrections.

Logarithmic corrections do in fact occur.  While in the abelian theory
the electrical conductivity only reaches a long wave length limit at a
scale set by a hard particle's free path for large angle scattering, in
the nonabelian theory that limit is set by the free path for {\em any}
scattering.  The reason is that when a colored particle undergoes a
scattering, however small the transfer momentum, its color is changed.
This degrades the color current even if it does not degrade the momentum
carried by the particle substantially.  The total rate for any
scattering to occur is twice the damping rate, which has been
computed for hard particles at leading log by Pisarski.  The damping
rate for an adjoint charged particle of any spin is \cite{Pisarski}
\beq
\gamma = \frac{N g^2 T}{4 \pi} \left[ \log \frac{m_D}{g^2 T} + 
	O(1) \right] \, ,
\label{value_gamma}
\eeq
and for a fundamental representation particle it is the same with $N
\rightarrow (N^2-1)/2N$.
Note the log, which arises from an integral over exchange momenta,
running from the $gT$ to the $g^2 T$ scale; also note that the result is
independent of the particle's momentum, provided it is harder than the
$gT$ scale.  A collision largely but incompletely 
randomizes a particle's charge,
and so on scales longer than $\sim 1 / ( g^2 T \log(1/g))$ the
electrical conductivity of the plasma is $\sim m_D^2 / 3 \gamma$, up to
nonlogarithmic corrections.  To find the numerical constant one
must determine how thoroughly a scattering randomizes a particle's
charge, which depends on the representation of the particle.  
Also, the particle's charge is not destroyed, just transferred to
another particle; one must check whether this induces any important
currents.  (It turns out not to, because a particle is as likely to
scatter from a charge carrier moving in one direction as in the exact
opposite direction.)  The calculation is quite nontrivial but it has
been done recently by Arnold, Son, and Yaffe \cite{ASY2}, who show that,
at leading log, the conductivity is simply $m_D^2 / 3 \gamma$, with
$\gamma$ given in Eq. (\ref{value_gamma}).  This is independent of the
group representation of the particles carrying the current.

Performing
an extreme parametric expansion, $\log (1/g) \gg 1$, the scale set by
particle mean free paths and the nonperturbative scale are well
separated, and the fields with $k \sim g^2 T$ see a $k$ independent
conductivity.  Hence the relevant
infrared dynamics for transverse modes is, at leading logarithmic order, 
\beq
[D_0 , A_i] = \frac{3 N g^2 T \log(1/g)}{4 \pi m_D^2} D_j F_{ji} + 
	{\rm noise} \, ,
\eeq
where the transverse part of the noise is fixed by the requirement that
the thermodynamics come out right.  The longitudinal part of the noise
generates time dependent gauge rotations of the $A$ fields, which are
irrelevant to Chern-Simons number; so we may choose the amplitude of the
longitudinal part of the noise to be whatever we want.  It is most
convenient to choose it to be of the same magnitude as the transverse
part, in which case we reproduce the effective theory of B\"{o}deker,
which is also the Langevin equation for 3-D Yang-Mills theory.

While this derivation has presented the ideas in an intuitive way it is
scarcely rigorous, so I will now approach the problem a little more
formally by looking at the hard thermal loop effective theory.

\begin{figure}
\centerline{\epsfxsize=6in\epsfbox{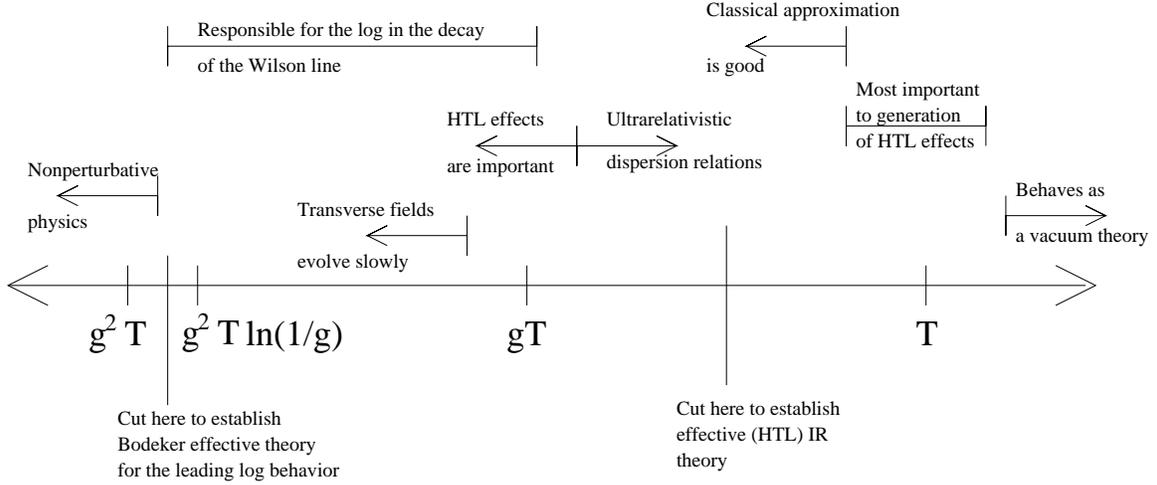}}
\caption{``It's quite simple, really $\ldots \;$.''  A scorecard of the
scales involved in the problem and the approximations which are valid in
each.  None of these scales are distinct if we do not take $g \ll 1$.}
\end{figure}

\subsection{HTL effective theory and the Wilson line}
\label{HTL_Wilson_subsec}

Now I will go through the argument for B\"{o}deker's effective
theory also from the point of view of the effective HTL theory for the
infrared modes.  The idea of the HTL effective infrared theory is that
one can construct an effective theory for the modes with $k \ll T$, valid 
at leading order in $g$, by integrating out all degrees of freedom with
$k \sim T$.  The procedure for separating the degrees of freedom and
regulating the effective theory, cut off at some scale between $gT$ and
$T$, is left unspecified and does not affect the result at leading order.  

The resulting effective theory is a classical theory for the remaining
modes, but with a nonlocal ``HTL'' effective action correction.  The
nonlocal effective action was first derived in \cite{HTLaction}.  In the
current context, since in the long term 
we have in mind a position space regulation of
the IR theory, it is most convenient to write the effective action in
real space, as was first done by Huet and Son \cite{HuetSon}:
\beqa
-[D_0 , E_i]^a(x,t) + [D_j , F_{ji}]^a(x,t) & = & 
	\xi_i^a(x,t) + \frac{m_D^2}{4 \pi}
	 \int d^3 y \frac{y_i y_j}{y^4}  
	U^{ab}((x,t),(x+y,t-y)) \times
	\nonumber \\ & & \qquad \qquad \qquad  \qquad E_j^b(x+y,t-y) 
	\, .
\label{HTL_realspace}
\eeqa
Here $U^{ab}((x,t),(x+y,t-y))$ is the adjoint parallel transporter along
the straight, lightlike path between the points $(x,t)$ and
$(x+y,t-y)$.  Note that the electric field on the right hand side is at
the retarded time $t-y$.  The noise $\xi_i^a$ is Gaussian with a
nonvanishing two point function for lightlike separated points,
\beq
\langle \xi^a_i(x,t) \xi^b_j(x+y,t') \rangle = \frac{m_D^2 T}{4 \pi} 
	\delta(|t-t'| - |y|)
	\frac{y_i y_j}{y^4} U^{ab}((x,t),(x+y,t')) \, .
\eeq

For $k \ll gT$, the effective action simplifies somewhat.
The coefficient of the term on the right in Eq. (\ref{HTL_realspace}),
which contains 
one time derivative (since $E_i = [D_0 , A_i]$), is large enough that
the time evolution is determined by this term and we can neglect the
$[D_0 , E]$ term on the left.  Further, the time scale associated with
the fields' evolution is longer than the time retardation appearing in
the nonlocal HTL action, and we are justified to neglect that
retardation.  The result is
\beq
[D_j , F_{ji}]^a (x) =  \xi_i^a(x) + \frac{m_D^2}{4 \pi}
	\int d^3 y \frac{y_i y_j}{y^4}  
	U^{ab}(x,x+y) E_j^b(x+y) \, ,
\label{simplified}
\eeq
and the noise correlator is simplified by dropping the retardation there
as well (though to get its magnitude right we must remember that there 
are two contributions from $\delta(|t-t'|-|y|)$).  If in addition 
$k \gg g^2 T \log(1/g)$ then it is possible to choose
a gauge such that the parallel transporter is close to the identity and
at leading order it can be ignored.  In this case we can recover
Eq. (\ref{sigma_intermed}) by expanding $[D_j , F_{ji}]$ to leading
order in $A$ and Fourier transforming.

Huet and Son argue that, to model the $O(g^2 T)$ modes alone, we can use
Eq. (\ref{simplified}), interpreted as an expression for fields at the
$g^2 T$ scale alone with all shorter wave length scales integrated out.  
The reasoning is that the higher modes only appear in the equation in
$[D_j , F_{ji}]$.  But $[D_j , F_{ji}]$ is a thermodynamical quantity
depending only on the transverse fields, and as already stated, the UV
causes perturbatively small corrections to this sector, which can be
ignored at leading order.  The power counting arguments are laid out
explicitly in \cite{Son}. 

However, this misses one key issue.  Can we
neglect the influence of the modes with $g^2 T \ll k \ll gT$ on the
adjoint parallel transporter $U^{ab}(x,x+y)$ when $y \sim 1/(g^2 T)$?
The answer is, no.  When evaluating an adjoint parallel transporter for
a path of length $\sim 1 / (g^2 T)$, essential contributions arise from
all scales intermediate between $gT$ and $g^2 T$.  This problem has
recently been addressed by Arnold and Yaffe, in the context of studying
$O(g^2 T)$ corrections to the Debye screening length
\cite{ArnoldYaffe}.  They show that, in SU($N$) pure gauge theory,
{\em any} two point correlator of
equal time adjoint fields at separation $y$, connected by a straight
adjoint Wilson line, falls off at least as rapidly as 
\beq
\exp(- y / \lambda ) \, , \qquad
\lambda^{-1} = \frac{N g^2 T}{4 \pi} \left( \log \frac{m_{\rm reg}}{g^2 T} 
	+ K \right) \,	, 
\label{HTL_reduced}
\eeq
where $m_{\rm reg}$ is a UV regulator and $K$ is a constant of order
unity.  The constant is evaluated for SU(2) 
in \cite{LainePhilipsen}; for the regulation appropriate to determining
the Debye mass, $m_{\rm reg} = m_D$ and $K = 6.7 \pm .3$.
Nonlogarithmic corrections are large in this case; I will return to
this point later.  The value of the constant $K$ includes
nonperturbative physics at the $g^2 T$ scale, but the logarithm arises
perturbatively from scales intermediate between $g^2 T$ and $m_{\rm
reg}$. 	Thus, to study physics on the length scale $1 / g^2
T$, we must include the influence of scales between $gT$ and $g^2 T$ on
the parallel transporter in Eq. (\ref{HTL_reduced}).

Note the sneaky way the modes with $g^2 T \ll k \ll gT$ have entered the
dynamics of the $g^2 T$ modes.  Their direct influence on the
interactions between the $g^2T$ modes is indeed small, as Son showed
\cite{Son}; but they change the way that the $g^2 T$ modes interact with
the hard modes, in a way which turns out to be important.  B\"{o}deker
emphasizes this viewpoint in his original derivation of the effective
theory for the leading log \cite{Bodeklog}.

Using the correction to the Wilson line from the intermediate momentum
modes, I can now establish B\"{o}deker's effective theory
for the $k \sim g^2 T$ modes.  The modes with $k \gg g^2 T$
change quickly compared to the $g^2 T$ scale modes, so the $g^2 T$
fields see an average over the more UV scales.  In particular the
parallel transporter relevant for the $g^2
T$ modes' evolution is the average of the parallel transporter
over realizations of the $ g^2 T \ll k \ll gT$
fields.  As I demonstrate in subsection \ref{exp_decaysec}, averaging
over realizations of $k \gg g^2 T$ modes leads to exponential damping of
the parallel transporter, for $y$ of order $1 / ( g^2 T \log(m_D/g^2
T))$.  In the
Coulomb gauge,\footnote{The use of Coulomb gauge becomes problematic
when considering length scales $l \geq 1/g^2 T$, and for considering any
unequal time correlator if the total volume of space considered is $V
\gg (g^2 T)^{-3}$; however this is not relevant because technically we
are only applying Coulomb gauge to modes with $k \gg g^2 T$ in order to
integrate them out and establish an effective theory.  The gauge fixing
of the IR effective theory, ie of the problematic $k \sim g^2 T$ modes, 
has not been specified.  Also, at
leading log the same results would be obtained in Landau gauge.}
\beq
\label{UCoulomb}
U_{\rm Coulomb} ^{ab}(x,x+y) \simeq \delta^{ab} \exp( - y / \lambda ) \, ,
\eeq
with $\lambda$ the same as in Eq. (\ref{HTL_reduced}).  If we are
permitted to expand in $\log(m_D/g^2 T) \gg 1$, then the integrand on
the right hand side of Eq. (\ref{simplified}) has already fallen away
before $y$ comes on order $1/(g^2 T)$; therefore the approximation which
gives Eq. (\ref{UCoulomb}) is valid, up to log corrections, throughout
the range of $y$ which dominates the contribution to 
the integral.  The effective theory for the $g^2 T$ modes
(fixing the gauge freedom on scales more UV than $g^2 T$ to Coulomb
gauge) is therefore
\beq
\xi_i^a + 
	\int d^3y e^{-(y/\lambda)} \frac{y_i y_j}{y^4}
	E^a_j(x+y) = [D_j , F_{ji}]^a(x) \, .
\eeq

Since the integral is dominated by $y \sim 1/(g^2 T \log(m_D/g^2T))$,
and for the IR fields of interest $E$ varies only on the $1/g^2 T$
scale, it is permissible at leading log order to pull the $E$ field out
of the integral.  The integral is then quite simple; performing it gives 
\begin{equation}
[D_j , F_{ji}] = \frac{m_D^2 \lambda}{3} E^a_i + \xi_i^a \, , \qquad
\langle \xi_i^a(x,t) \xi_j^b(y,t') \rangle = \frac{2 Tm_D^2 \lambda}{3} 
	\delta(x-y) \delta(t-t') \delta_{ij} \delta_{ab} \, ,
\label{Effect_theory}
\eeq
where the form for the noise correlator also follows from the
approximation for the parallel transporter.  (Alternately, one can always
recover the form of $\xi$ by insisting that the thermodynamics come out
correctly. )
This is B\"{o}deker's effective theory, though it remains to establish
that $m_{\rm reg}$ should be $m_D$.  I discuss this more in the next
section.  

\section{Wilson line, more carefully}
\label{Wilsonline}

Now it is time to look more carefully at the Wilson line appearing in
the last section, first to verify the claims there, second to show the
connection to the argument involving scatterings of the hard particles,
and third because it is relevant to the analysis of the results of Hu,
M\"{u}ller, and Moore.

\subsection{Exponential decay of the Wilson line}
\label{exp_decaysec}

What we want to know about is the Wilson line between lightlike
separated points a spatial distance $l$ apart, with $l \gg 1 / m_D$ but
less than $1 / (g^2 T)$ by at least a logarithmic factor.  Actually, we
want to know the average of the Wilson line over realizations of the
modes with $k \gsim g^2 T \log (1/g)$, since these fields fluctuate
faster than the $O(g^2 T)$ fields, which therefore see the average over
realizations, up to corrections subdominant in $1 / \log (1/g)$.

The Wilson line is given by 
\begin{equation}
U = P \exp \int_0^l i g (A_0 + A_z)(z,t=z) dz \, ,
\end{equation}
with $A$ in the adjoint representation.  I will only consider the
transverse contributions here, the longitudinal ones are subdominant.  I
use Coulomb gauge, in which the $A_0$ contribution arises entirely from
the longitudinal modes and the $A_z$ only arises from the transverse
modes.  Further, I will work at leading parametric
order, by which I mean that higher point correlators and vertex
insertions are ignored, and combinations of $A$ fields are evaluated
assuming $A$ is Gaussian by applying Wick's theorem.  Of course I will
include the hard thermal loop corrections to the $A$ field propagators.
These approximations are justified at leading log down to $k \sim g^2 T
\log(1/g)$, which is all we need.  (If I were interested in lengths $l
\sim 1/g^2 T$ rather than logarithmically shorter, most of the
approximations I make would break down completely.)

All terms with odd powers of $A$ vanish on averaging over realizations,
while even terms look like (applying Wick's theorem between lines 1 and
2) 
\beqa
\langle U \rangle & = & \sum_{n=0}^{\infty}
	\frac{(-)^n g^{2n}}{(2n)!} \int dz_1 \ldots dz_{2n} \langle 
	A_z(z_1) \ldots A_z(z_{2n}) \rangle \\
	& = & \sum_{n=0}^{\infty} 	
	\frac{1}{n!} \left( \frac{-g^2}{2} \int_0^l dz_1 dz_2 \langle 
	A_z(z_1) A_z(z_2) \rangle \right)^n \nonumber \\
	& = & \exp \left(  \frac{-g^2}{2} \int_0^l dz_1 dz_2 \langle 
	A_z(z_1) A_z(z_2) \rangle \right) \nonumber \, ,
\eeqa
so the average over realizations is the identity times the exponential
of the two point contribution.  (Note that for any given realization the
Wilson line has unit modulus.  But the average over realizations does not,
its modulus falls exponentially with distance.)

To evaluate this we need the two point correlator,
\beq
\langle A_i(x,t) A_j(y,t') \rangle = \int \frac{d^3 k}
	{(2 \pi)^3} \frac{d \omega}{2 \pi} \frac{T}{\omega} 
	\rho(k , \omega) \left( \delta_{ij} 
	- \frac{k_i k_j}{k^2} \right) e^{i\omega(t'-t)} e^{-ik 
	\cdot (y-x)} \, .
\eeq
Here $\rho(k,\omega)$ is the spectral density, which is the magnitude of
the discontinuity in the propagator $1 / ( \omega^2 - k^2 - \Pi_T(k ,
\omega))$ across the real $\omega$ axis on analytic continuation from 
Euclidean (imaginary) $\omega$, 
\beq
\rho(k , \omega ) = 2 {\rm Im} \left( (\omega + i
	\epsilon)^2 - k^2 - \Pi_T(k , \omega+i \epsilon) \right)^{-1} \, ,
\eeq
and $T / \omega$ is the classical approximation for $1 + n(\omega)$,
with $n(\omega)$ the Bose distribution function.

So far I have suppressed group indices, but
when they and the integrals over $z$ are evaluated we get 
\beq
\log \frac{ {\rm Tr} \: \langle U \rangle}{{\rm Tr} \: {\bf 1}} = 
 - \frac{N g^2}{2} \int \frac{d^3 k}{(2 \pi)^3} \frac{ d\omega}{2 \pi} 
	\frac{T \rho(k,\omega)}{\omega}
	\frac{k_x^2 + k_y^2}{k^2} \, \frac{ 4\sin^2 
	(k_z - \omega) l/2}{(k_z - \omega) ^2} \, .
\label{exp_decay}
\eeq

The equal time correlator of the $A$ field goes as $T / k^2$
plus subleading corrections, for all $k \gg g^2 T$; so
\beq
\int \frac{d \omega}{2 \pi} \frac{T \rho(k , \omega)}{\omega} 
	= \frac{T}{k^2} \, .
\label{total_spect_wt}
\eeq
What matters now is where $\rho$ is concentrated.  In the regime 
$k \ll m_D$, almost all the contribution to Eq. (\ref{total_spect_wt})
is from $|\omega| \ll k$, see \cite{Pisarski}.  This is just the statement
that these modes evolve on time scales slower than $1/k$, as I have
already discussed.  Hence, in evaluating the low $k$ contribution to 
Eq. (\ref{exp_decay}) I can set $k_z - \omega \simeq k_z$, and then
perform the integral over $\omega$, giving
\beq
 - \frac{N g^2 T}{2} \int^{|k| \leq m_D} \frac{d^3 k}{(2 \pi)^3} \,
	\frac{1}{k^2} \, 
	\frac{k_x^2 + k_y^2}{k^2} \, \frac{ 4\sin^2 
	k_z l/2}{k_z^2} \, .
\eeq

The $k_z$ integral is completely well behaved; in fact the large $l$
limit of $4 \sin^2(k_z l/2) / k_z^2$ is $2 \pi l \delta(k_z)$.
This represents the fact that only modes with $k_z < 1/l$ have $A$ of
the same phase all along the Wilson line; other modes' contributions
destructively interfere in the integral along the Wilson line.  The
integral over the other two directions is dominated by a logarithm
arising from scales intermediate between $k_\perp \sim 1/l$ and $k_\perp
= m_D$, where the approximation for $\rho$ breaks down.
Performing the integral over $k_\perp$ first gives
\beq
- \frac{N g^2 l T}{4 \pi^2}\left( \log \frac{m_D}{l}+O(1) \right) 
	\int \frac{ \sin^2 (k_zl/2)}{(k_zl/2)^2 } d(k_zl/2) = 
	- \frac{N g^2 l T}{4 \pi}\left( 
	\log \frac{m_D}{l}+O(1) \right) \, .
\eeq

In the opposite limit, $k \gg m_D$, the excitations obey normal vacuum
ultrarelativistic dispersion relations to good approximation, so the
spectral density is approximately
\beq
\rho(k \gg m_D,\omega) \simeq \frac{\pi}{k} ( \delta(k - \omega) -  
	\delta(k + \omega)) \, .
\eeq
The large $k$ contribution is then
\beq
- \frac{N g^2}{2} \int_{m_D} \frac{d^3 k}{(2 \pi)^3} \,
	\frac{k_x^2 + k_y^2}{k^2} \, \frac{T}{k^2} \, \frac{ 4\sin^2 
	((|k|-k_z) l/2}{(|k| - k_z)^2} \, .
\eeq
Now the term with $\sin^2$ in it is forcing $|k| = k_z$; only modes
propagating along the direction of the Wilson line keep in phase, others
destructively interfere.  But the polarization vector of such a mode is
close to orthogonal to the Wilson line; because $k \gg 1/l$, for any $k$
for which $|k|- k_z < 1/l$, the factor $k_\perp^2 / k^2$ will be near
zero.  Thus, the only modes which avoid 
destructive phase interference are polarized in the wrong direction to
contribute significantly.  Continuing to carry out the integral by
defining $x = k_z / k$, we get
\beq
-\frac{N g^2 T}{2 \pi^2} \int_{-1}^{1} dx \int_{m_D} dk (1-x^2) 
	\frac{4 \sin^2 (lk(1-x)/2)}{(1-x)^2} \sim -\frac{N g^2 T}{2
	\pi^2 m_D} \left( \log (l m_D) + O(1) \right) \, ,
\eeq
which is $O(g)$.  The hard modes do not contribute at leading order to
the Wilson line along a lightlike path.

I have not treated the modes with $k \sim m_D$, which are more
complicated because this is where $\rho$ does not fit into either
limiting category.  But they turn out to give a result smaller by a
logarithm than that from the modes with $k \ll m_D$.
The final result is that, on averaging over realizations of the
modes with $k$ greater than $g^2 T$ by at least a logarithm, the Wilson
line in Coulomb gauge is
\beq
\langle U \rangle = {\bf 1} \exp( - l / \lambda ) \, , \qquad
	\lambda^{-1} = \frac{N g^2 T}{4 \pi} \left( \log \frac{m_D}{g^2 T} 
	+ O(1) \right) \, .
\eeq

\subsection{relation to scattering}

Now look at Eq. (\ref{exp_decay}) again.  Remember that the Wilson line
is representing the trajectory of a hard particle, with $p$ much greater
than any momentum scale which gives a leading order contribution to the
integral.  We want to interpret Eq. (\ref{exp_decay}) as $l$
times the rate for the particle to undergo a scattering involving the
transfer of a soft field, times a (representation dependent) group theory
factor which tells how thoroughly the scattering randomizes the particle
charge.  To see the relation, take the large $l$ limit.
Then $4 \sin^2 ((k_z-\omega)l/2) / (k_z-\omega)^2 = 2 \pi l \delta(k_z -
\omega)$, and the decay rate per unit length of the Wilson line is
\beq
\frac{N g^2}{2} \int \frac{d^3 k}{(2 \pi)^3} \, \frac{d\omega}{2\pi} \, 
	\frac{T \rho(k,\omega)}{\omega} \,
	\frac{k_x^2 + k_y^2}{k^2} \, 2 \pi \delta ( k_z - \omega ) \, .
\label{scatter_rate}
\eeq
Recall that the hard particle starts out with $p_z \gg
|k|$, $p_x=p_y=0$.  If it emitted a particle of wave number $k$, its
momentum would change to $\vec{p} - \vec{k}$ and its energy would change
by $-k \cdot \hat{p} = -k_z$ plus a correction of order $k^2 / |p|$, which
is negligible by assumption.  The delta function is just the energy
conserving delta function appearing in the expression for the rate of
the process shown in Figure \ref{scatter} $(a)$, the emission of a soft
gluon by an adjoint charged, hard mode.  

\begin{figure}[t]
\centerline{\epsfxsize=6in\epsfbox{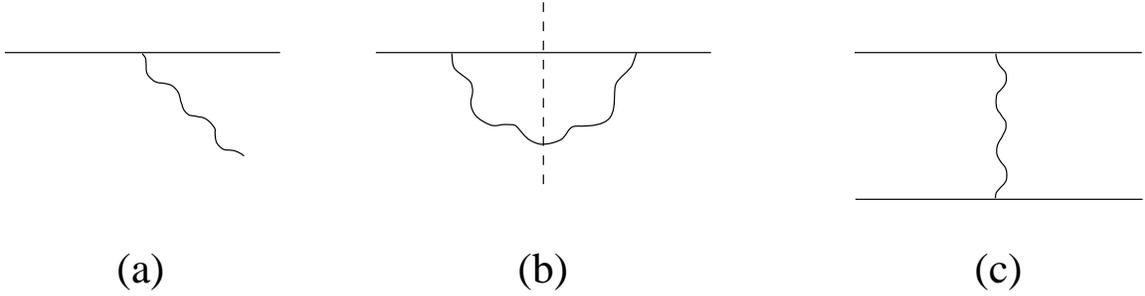}}
\caption{\label{scatter} (a) Emission of a gluon below the light cone;
(b) Self-energy diagram which, cut, describes the emission; (c) The cut
must split an HTL insertion, so the physical process is scattering with
a small exchange momentum.}
\end{figure}

The rate for this process has
been considered by Pisarski \cite{Pisarski}; his expression (4.4) does
not look quite the same as Eq. (\ref{scatter_rate}) for the
following reasons.  First, his definition of the spectral density
differs from the one used here by a factor of $2 \pi$.  Second, he has
allowed the hard particle to be a little off shell, which is important
in the part of the integral with $k \sim g^2 T \log(1/g)$ but not for
higher $k$. (Note however that my large $l$ approximation breaks down
in exactly this regime.)
For momenta $k \gg g^2 T \log(1/g)$ we may integrate over $\omega_t$ in
his expression and obtain Eq. (\ref{scatter_rate}).  The difference
caused by $k \sim g^2 T \log(1/g)$ does not change the leading log, but
would be important in investigating corrections to the leading log.
Also note that Pisarski considers the contribution of longitudinal
momenta and finds that they do not give a log, only a constant times $N
g^2 T$.

The process considered involves the emission of a soft mode with $|k| > 
|\omega|$.  The reason that such emission is possible at all is because
of the hard thermal loop correction to the gauge propagator.  The rate
is the imaginary part obtained by cutting the one loop self-energy
diagram, Fig. \ref{scatter}$(b)$.  The cut gives a nonzero result
when it goes through a HTL self-energy insertion, which can be
interpreted as diagram $(c)$ in the figure.  Hence the Wilson loop
calculation is proportional to the rate for scatterings of one hard
particle off all other hard particles by exchange of soft intermediaries.

Pisarski calculates the total rate for a particle to undergo a
collision, and because of the conventional definition of the damping
rate, his damping rate is actually half of the scattering rate.  Also,
his result does depend on the representation of the particle
undergoing the scattering.  The rate the Wilson line calculation
determines is actually the rate of color randomization, not of
collisions; there is a representation dependent correction between the
two, which depends purely on color factors at the vertex where the
particle of interest interacts with the soft background field.  
The total collision rate of a particle is proportional to the group
factor 
\beq
\frac{{\rm Tr} T^a T^a}{{\rm Tr} {\bf 1}} \, ,
\eeq
where $T^a$ is in whatever representation the particle is in.  The
original disturbance of the particle distribution from equilibrium is
caused by an electric field,
which is an adjoint object; the disturbance to the single
particle density matrix is proportional to $E^b T^b$.  The color
randomization per collision is
\beq
1 - \frac{{\rm Tr} {\bf 1}}{{\rm Tr} T^a T^a} \, \frac{T^a T^b T^a}{T^b}
	\, ,
\eeq
where the second term tells how much the color after the collision is
aligned with the color before.  Multiplying
by the total collision rate gives a color
randomization rate $\propto N/2$, independent of representation.

\subsection{subleading corrections in the method of Hu and M\"{u}ller}
\label{HuMuller_subsec}

Now I will discuss the relation between the quantum theory and 
the technique proposed by Hu and M\"{u}ller, refined and implemented
jointly with myself, concentrating on whether the behavior is the same
at next to leading logarithmic order.

First I should explain why the technique is necessary.  Traditionally
people have tried to determine the sphaleron rate by studying classical
Yang-Mills (or Yang-Mills Higgs) theory regulated on a spatial lattice.
However, if we just
study the classical theory on the lattice, using the lattice spacing as
the UV cutoff, the hard lattice modes generate HTL effects which do not
look like Eq. (\ref{HTL_realspace}) and in fact are not rotationally
invariant \cite{Smilga}.  Even in the leading log approximation this is
a problem, because it means that, where Eq. (\ref{Effect_theory}) has
$E_i$, we will get $E$ rescaled by a rotationally non-invariant factor,
determined in \cite{Arnoldlatt}.

A proposal by Arnold 
to fix this problem, staying within lattice classical theory,
by making the dispersion relations turn up very steeply
\cite{Arnoldlatt}, does not work because the hard modes are then Landau
damped\footnote{The problem with Arnold's proposal is that it gives a
Lorentz non-invariant hard mode dispersion relation under which
$1\rightarrow 2$ and $2 \rightarrow 1$ processes are kinematically
allowed.  They are efficient, so the hard excitations have a mean free
path for hard scatterings $\sim 1 / g^2 T$.  I have Arnold's
agreement on this point.}.  
The only alternatives I am aware of involve
adding new degrees of freedom which influence the IR classical fields in
a way equivalent to correct hard thermal loops.  Two such proposals
exist in the literature.  One is due to B\"{o}deker, McLerran, and Smilga
\cite{Smilga}, more recently discussed by Iancu \cite{Iancu}.  I will
not discuss it since no one has yet specified a complete discrete numerical
implementation.  The other idea was proposed by Hu and M\"{u}ller
\cite{HuMuller}; the details of the implementation were worked out and
applied jointly with me \cite{particles}.

We also simulated the classical IR physics by studying nonperturbatively
the classical system regulated on a spatial lattice, thereby treating
the left hand side of Eq. (\ref{HTL_realspace}) 
fully nonperturbatively.  To include the HTL effective action, the right
hand side of the expression, we added to the classical lattice system a
large number of adjoint charged classical particles.  They take
coordinate positions in the continuous space in which the lattice fields
sit, obey ultrarelativistic dispersion relations, and interact with the
lattice fields when they cross the dual planes to lattice links.  There
are two parts to the interaction with the lattice fields.  

First, the
particles ``kick'' lattice electric fields, and their momenta receive a
similar ``kick''.  The kicks and the approximately random distribution
of the charges performs the noise, and a correlation between past fields
and the ``kick'' the gauge field receives, arising from a change in the
particle trajectory from the ``kick'' it received, accounts for the
nonlocal term.  The size of each kick is 
proportional to a charge $Q$ which is
made small so the particles individually interact weakly.  

\begin{figure}[t]
\centerline{\epsfxsize=3in\epsfbox{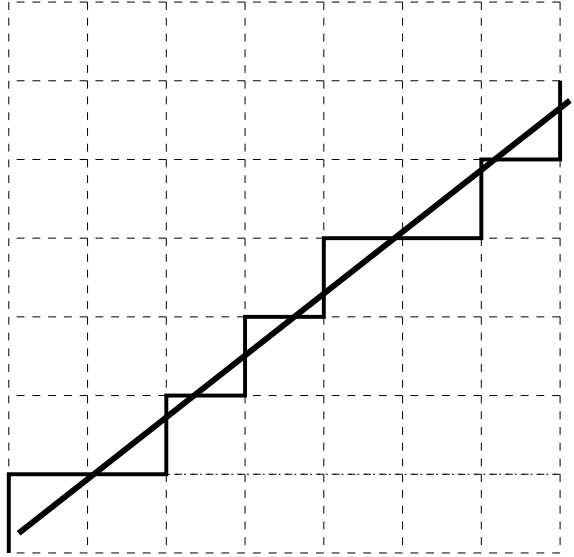}}
\caption{\label{fig2} A simple 2-D example of how a Wilson line (diagonal)
on a lattice (dashed lines) is replaced by the sequence of links which
stay closest to it (solid, jagged line).}
\end{figure}

Second, the
particles' adjoint charges are parallel transported by the gauge field
connection.  Since the gauge fields are only defined on the lattice
links, the Wilson line used for the parallel transportation of a
particle is the sequence of lattice links which maintains the shortest
distance to the actual path of the particle, as illustrated in Figure
\ref{fig2}.  The exact rule used to choose the sequence of links is that
a link is used if the Wilson line penetrates the face of the dual lattice
corresponding to that link.  Parallel transporting the particle's charge
ensures that we are including the parallel transporter in the right hand
side of Eq. (\ref{HTL_realspace}); it is also absolutely necessary to
make the update rule manifestly gauge covariant.  The full update is
described in excruciating detail in \cite{particles}.

The Debye mass depends on the density $\langle n \rangle$ and charge $Q$
of the added particles as $m_D^2 \propto Q^2 \langle n \rangle$, 
so by changing the number of particles
one can tune the HTL strength to be whatever is needed while keeping $Q$
small.  If there were only IR fields, meaning fields with
$k \ll 1/a$ ($a$ the lattice spacing) then the behavior would correctly
reproduce Eq. (\ref{HTL_realspace}), at least in the limit $Q
\rightarrow 0$, $\langle n \rangle \rightarrow \infty$ with $m_D^2$
fixed.  This is also discussed in \cite{particles}.

The lattice theory involves two scales, $a$ and $m_D$; to get the
leading parametric behavior we must seek the limits $a \rightarrow 0$,
$m_D \rightarrow \infty$ (if we think of the length scale $1 / (g^2 T)$
as remaining fixed).  There are two ways we could go about doing
this.  One corresponds to using $a$ as a cutoff between the $gT$ and the
$T$ scales, meaning that we maintain $a \ll 1 / m_D$ but $a \gg g^2 T /
m_D^2$.  The latter condition is essential to make sure that ``bad'' 
hard thermal loops arising from modes with $k \sim 1/a$, which have the
wrong dependence on $k,\omega$ and in particular are not rotationally
invariant \cite{Smilga}, are subdominant
to ``good'' hard thermal loops due to the particles.  The other way of
taking the limits is to make $m_D \gg 1/a$, so the wave number $(1/a)$
falls between the $gT$ and $g^2 T$ scales.

In each approach, the model correctly generates the effective HTL
dynamics up to power corrections in $(g^2 a T)$, $(Q/g^2 a T)$, $g^2 T
/m_D$, and $(g^2 T)/(a m_D^2)$, except perhaps for the behavior of the 
parallel transporter.  The contribution to the parallel transporter from
IR fields is correct because for a gauge field of wave number $k \ll
1/a$ the replacement of
the straight line path with the ``jagged'' path actually used (see again
Figure \ref{fig2}) gives the right behavior up to corrections suppressed
by $O(k^2 a^2)$.  The question is, how does the UV contribute to the
parallel transporter?

In the case where we make $m_D \gg 1/a$, {\em all} of the lattice
degrees of freedom have overdamped evolution.  None of them propagate and
there are no problems from hard thermal loops arising from classical
lattice degrees of freedom with $k \sim 1/a$.  Except for the parallel
transporter, the systematics are then best under control.  However the
log appearing in the parallel transporter is obviously cut off by the
inverse lattice spacing, $\log (m_D / g^2 T)$ becomes $C + \log(1/g^2 a
T)$ with $C$ a coefficient to be determined by matching.  Since we know
that the lattice regulator is not rotationally invariant we expect that
$C$ will be direction dependent, leading to a rotationally non-invariant
correction.  However, this correction is only at next to leading order
in an expansion in $\log (1/g^2 aT)$.

In the other case, $m_D \ll 1/a$, all of the physics which sets both
upper and lower limits on the log in Eq. (\ref{exp_decay}) lie at $k \ll
1/a$, and naively we should get the correct behavior at leading order in
$g$.  We do not, however, because the most UV lattice modes have
dispersion relations which ``turn over,'' $k > \omega$.  Diagram $(a)$
from Figure \ref{scatter} can occur with the emitted gluon on shell.  In
fact this is just \v{C}erenkov radiation, which is permitted because the
group velocity of the most UV lattice modes is subluminal, while the
hard particles move at the speed of light.  Hence, while there is no
contribution to the parallel transporter from modes with 
$1/a \gg k \gg m_D$, there is a contribution from $k \sim 1/a$, and
again we get a rotationally non-invariant correction, $\log (m_D / g^2
T)$ becomes $C + \log(m_D / g^2 T)$.

To determine the value of $C$ we need to repeat the arguments leading to
Eq. (\ref{exp_decay}) but using lattice gauge fields multiplied along a
``jagged'' path which stays closest to the straight line path under
consideration.  For a Wilson line of length $l = L a$ in a direction
with unit vector $p_i$, with $p_i$ in the first octant,
the equivalent of Eq. (\ref{exp_decay}) is  
\beqa
\label{Wilson_latt}
\log \frac{ {\rm Tr} \langle U \rangle}{{\rm Tr} {\bf 1}} & \! \! = \! \! & 
	 - \frac{N g^2}{2} \int_{-\pi/a}^{\pi/a} 
	\frac{d^3 k}{(2 \pi)^3} \int_{-\infty}^{\infty} 
	\frac{ d\omega}{2 \pi} \frac{T \rho(k,\omega)}{\omega}
	F(k , \omega) \, , \\
F(k , \omega) & \! \! = \! \! & \sum_{\epsilon} 
	\int_{0}^{1} \! dx dy dz \left\{
	a \epsilon_x \! \! \sum_{n_x = 1}^{[L p_x + x]} \! \exp i \Bigg(
	\omega \frac{n_x-x}{p_x} - k_x (n_x-.5)
	- k_y \left[ y+ \frac{n_x-x}{p_x}p_y \right] - \right. 
	\nonumber \\ & & \qquad \qquad \qquad \qquad \left.
	- k_z \left[ z+ \frac{n_x-x}{p_x}p_z \right] \Bigg)
	+ \left( \sum_{n_y} \, , \:
	\sum_{n_z} \: {\rm similar} \right) \right\}^2 \, ,
\label{Wilson_latt2}
\eeqa
and the values in other octants follow from cubic symmetry.
Here the sum over $\epsilon$ is over a basis of the two unit vectors
satisfying $\sum_i \epsilon_i \sin(k_i a/2) = 0$.  The integral over
$(x,y,z)$ appearing in the definition of $F$ averages over starting
positions for the Wilson line within a lattice cell, and square brackets
always mean the argument is rounded down to an integer.  The rounding
down to an integer makes it very difficult to evaluate the expression
analytically.  It is also tricky to find $\rho(k,\omega)$ on the
lattice, away from the $m_D \gg 1/a$ or $m_D \ll 1/a$ limits.

In the infrared, meaning $k \ll 1/a$, the
sum in $F$ is well approximated by an integral, recovering
Eq. (\ref{exp_decay}).  In the UV, while Eq. (\ref{Wilson_latt}) 
is manifestly cubic symmetric,
it is not spherically symmetric:  the $k$ integral is over a cubic
range, and neither $F(k,\omega)$ nor $\rho(k,\omega)$ are rotationally
invariant.  The integral has been computed by Arnold
and Yaffe in the large $L$ limit for the special case that $p_i$ lies in
a lattice direction and the spectral density is concentrated at $\omega
= 0$ \cite{ArnoldYaffe}, although they did not present their calculation
in this context.  I will now compute it in the opposite limit, $m_D \ll
1/a$, but still along a lattice direction.  
If $L$ is large and $p_i$ lies in a lattice direction
(say, the $z$ direction) then $F(k, \omega) \simeq 2 \pi l \delta ( k_z
- \omega \; mod \; (2\pi/a))$. For $m_D \ll 1/a$ the
dispersion relation is the free lattice dispersion 
relation $\omega^2 a^2 = \sum_i ( 2 - 2 \cos(k_i a))$ and the spectral
density $\rho$ all lies on shell, and the integral reduces to 
\beq
\frac{Ng^2 lT}{2} \int_{- \pi/a}^{\pi/a} \frac{d^3 k}{(2 \pi)^3} 
	\frac{\tilde{k}_x^2 + \tilde{k}_y^2}{(\tilde{k}^2)^2}
	2 \pi \delta \left( k_z - \sqrt{\tilde{k}^2} \; {\rm mod} \: 
	\frac{2 \pi}{a} \right) \, ,
\eeq
where $\tilde{k}_i = 2 \sin(k_i a/2) / a$.  Evaluating numerically
gives $ .0466539 N g^2 l T$, which is $.586$ times the coefficient of
$\log (m_D / g^2 T)$.  This demonstrates that there
are extra UV contributions arising from $k \sim 1/a$ in the case $m_D \ll
1 / a$, which though not terribly large are not negligible.

The upshot is that however $m_D$ is tuned, the UV lattice degrees of
freedom will introduce a rotationally non-invariant contribution to the
radiative correction to the Wilson line which determines the parallel
transport of a particle's charge.  This spoils the rotational invariance
of the IR HTL effective theory at next to leading order in
$\log (1/g)$.  The best we could do would be to average the value of
Eq. (\ref{Wilson_latt}) over angles.  Then an appropriate
choice of lattice spacing $a$ in the regime with $m_D \gg 1/a$ could match
the value of the $\log( m_D / g^2 T)$ as closely as possible to the
quantum theory value.  This is the best option I see for determining the
sphaleron rate beyond leading order, but it does not eliminate all
systematic errors even at leading parametric order in $g$.

I have not shown that the same problem will arise for other possible
lattice implementations of HTL effects; but since such implementations
must generally involve parallel transportation on the lattice I expect
that the problem discussed here is general.  Certainly, when proposing
some other numerical implementation of HTL's, the burden of
proof must lie on the side of showing that problems from rotational
non-invariance of Wilson lines do not arise.

\section{Numerics}
\label{numerics}

Now that I have discussed the establishment of the effective theory,
Eq. (\ref{Effect_theory}), 
I will discuss how to make a lattice model of that
effective theory and how I compute the leading log coefficient of the
sphaleron rate by using that model.  First, define a Langevin time
$\tau$ related to the time appearing in Eq. (\ref{Effect_theory}) via
\beq
\label{tau_defined}
d\tau \equiv \frac{3}{ m_D^2 \lambda} dt = \frac{3N g^2 T \log(m_D/g^2T)}
	{4 \pi m_D^2} dt \, .
\eeq
Note that $\tau$ has dimensions of length squared, not length.
The effective infrared theory then has the familiar form
\beq
[D_\tau , A_i]^a (x,\tau) = [D_j , F_{ji}]^a(x,\tau) + \xi_i^a(x,\tau)
	\, , \qquad \langle \xi_i^a(x,\tau) \xi_j^b(y , \tau') \rangle = 
	2 T \delta(x-y) \delta(\tau - \tau') \, ,
\label{Langevin_eq}
\eeq
which is a Langevin equation.  Studying it, I will find the diffusion
constant for Chern-Simons number
\beq
\Gamma_{\rm Langevin} = \lim_{V \rightarrow \infty} \lim_{\tau 
	\rightarrow \infty} \frac{\langle (N_{\rm CS}(\tau) 
	- N_{\rm CS}(0) )^2 \rangle }{V \tau} \, ,
\eeq
numerically.  To do that I find the diffusion constant per lattice site
per $a^2$ of Langevin time, and multiply by $a^{-5}$.  On dimensional
grounds $\Gamma_{\rm Langevin}$ must be of order $\alpha_W^5 T^5$.   
Using Eq. (\ref{tau_defined}), the relation between
$\Gamma_{\rm Langevin}$ and $\kappa'$ defined in Eq. (\ref{def_kappa})
is 
\beq
\kappa' = \frac{3 N}{4 \pi} \frac{\Gamma_{\rm Langevin}}
	{\alpha_W^5 T^5} \, .
\eeq

Before getting further it is worth commenting that the IR behavior of 
Eq. (\ref{Langevin_eq}) is insensitive to the UV regulation and the
limit in which that regulation is removed, exists.  Though this
statement appears banal, it makes B\"{o}deker's effective theory completely
different from the classical Hamiltonian dynamics.  The essential
difference is that there are no long time scale correlations for the UV
fields in the Langevin evolution; a mode 
with wave number $k$ gets randomized in Langevin time $\tau \sim 1 /
k^2$, which is much faster than the natural time scale for the evolution
of IR fields.  Hence the IR fields see the average over all excitations
of the UV fields.  The influence of the UV must be purely thermodynamic,
and we know from the super-renormalizability of 3-D Yang-Mills theory
that the thermodynamic influence of the UV on the transverse sector is
well behaved.  For a more rigorous presentation of the argument see
\cite{ASY2}, who show that any purely dissipative update algorithm will
give a good continuum limit.  
For the Hamiltonian system, on the other 
hand, UV modes are propagating; the unequal time correlator behaves like
$\cos(kt)$ rather than like $\exp(- k^2 \tau)$.  The existence of long
time scale correlations of the UV fields is what makes the HTL effects
important to the IR dynamics.  Because of this difference, we can expect
a good small lattice spacing $a$ limit to exist for the Langevin time
dynamics, and it is worth it to try to control systematic errors.

I discuss the continuum $\tau$, spatial lattice implementation of 
Eq. (\ref{Langevin_eq}) in the
appendix; here I will just mention how I discretize the time update.
I define a time step $\Delta \tau = a^2 \Delta$, $\Delta$ a pure number
much less than 1.  The fields will be well defined at times $n \Delta
\tau$, $n$ an integer.  The noise is constant in each interval $[n
\Delta \tau, (n+1)\Delta \tau)$; its value at each point, direction,
and Lie algebra direction is drawn from the Gaussian
distribution with mean value $\sqrt{2T / a^3 \Delta \tau}$; its value at
each point, direction, and Lie algebra direction, and in
each time interval, is independent.  To determine the fields at time $(n+1)
\Delta \tau$ from the fields at time $n \Delta \tau$ I use the following
second order algorithm:  I compute $dA/d\tau(n \Delta \tau + 0)$ and
use it to predict $A((n+1) \Delta \tau)$;  then I average the values of 
$dA/d\tau$ at the starting point and at the predicted end point, and use
this average to update 
$A(n \Delta \tau)$ to time $(n+1)\Delta \tau$.  The step size errors are
$O(\Delta^2)$.  In particular, if I were studying the free theory a mode
with wave number $k$ would be updated with step size errors $\simeq (k^2
\Delta \tau/2)^2$.  The direct errors in the update of the IR fields are
tiny.  However, the most UV modes are only updated correctly if $\Delta \ll
1$.  The UV fields influence the IR modes radiatively, so we do need
$\Delta$ to be small; but the radiative corrections are suppressed by
$O(g^2 aT)$, and it is not too difficult to get $\Delta$ small enough to
make step size errors subdominant to statistics.

Unfortunately the above update is quite inefficient.  However there is a
much more efficient algorithm for dissipatively updating the fields,
the heat bath algorithm.  Rather than applying a very
small step of Langevin update to each lattice link in parallel, the idea
is to go through the links of the lattice at random, performing a
complete heat bath update of each link.  The
relation between Langevin time and the number of links updated is
discussed in Appendix \ref{AppendixA}; in particular it is possible to
make a very accurate match between the Langevin and heat bath time
scales by measuring the autocorrelations of some IR observable.  I apply
the small $\Delta$ limit as part of the matching, so that is taken care
of.  

Appendix \ref{AppendixA}, together with previous work \cite{Oapaper},
shows how to control lattice spacing systematics so they 
first appear at $O(a^2)$.  Also, it is possible to define a lattice
measurable to use for $N_{\rm CS}$ which is topological and will
eliminate systematic errors in $\Gamma$ due to the definition of $N_{\rm
CS}$.  In fact, two fairly efficient techniques are available
\cite{slavepaper,broken_nonpert}; 
here I will use the method developed in \cite{broken_nonpert}.
It remains to take the large volume and time limits.  
It was shown in \cite{AmbKras} that finite volume systematics are
negligible on cubic toroidal lattices larger than $8 / g^2 T$ on a
side.  To be doubly sure, I have used a lattice $16 / g^2 T$ on a side;
as a check I measure $\Gamma$ also on a lattice of half this size to
check that the result is the same.  Taking the infinite time limit is
tied up with the problem of converting a Langevin time series for
$N_{\rm CS}$ into a measurement of $\Gamma$.  I use the same analysis
techniques as \cite{slavepaper}.

\begin{table}
{\centerline{\mbox{
\begin{tabular}{|c|c|c|c|} \hline
lattice spacing $a$ & Volume & Langevin time &
	$\kappa' \pm$ statistical error \\ \hline
$2/3g^2T$  &  $(  8 / g^2 T)^3$ & $290000a^2$ & $10.44 \pm 0.23$ \\ \hline
$2/3g^2T$  &  $( 16 / g^2 T)^3$ & $49500 a^2$ & $10.30 \pm 0.21$ \\ \hline
$2/5g^2T$  &  $( 16 / g^2 T)^3$ & $21000 a^2$ & $10.70 \pm 0.67$ \\ \hline
$2/7g^2T$  &  $( 16 / g^2 T)^3$ & $42000 a^2$ & $10.26 \pm 0.79$ \\ \hline
\end{tabular}}}}
\caption{\label{thetable}
	Results for $\kappa'$ at two lattice spacings and two lattice
volumes.  The results show excellent spacing and volume independence.}
\end{table}

To verify good control of lattice spacing systematics I have made
measurements of $\Gamma$ at three lattice spacings, $a = 2 / 3 g^2 T$
($\beta_L = 6$), $a = 2/5 g^2 T$ ($\beta = 10$), 
and $a = 2 / 7 g^2 T$ ($\beta_L = 14$).
The results are presented in Table \ref{thetable}.  Finite volume and
spacing systematics are under control.  In particular, varying the
lattice spacing by over a factor of two leads to corrections smaller
than the statistical errors.  This makes large $a$ extrapolation
unnecessary, which is very important, since numerical cost
rises as $(1/a)^5$.

If I had used naive rather
than radiatively corrected relations between lattice and continuum
parameters, then the value of $\kappa'$ at the three lattice
spacings would be $17.7$, $15.0$, and $13.3$ respectively.  
Such strong dependence is
because converting $\Gamma$ from lattice to continuum units involves the
5'th power of $a$, and the radiative corrections are at leading order a
shift in the meaning of $a$ from the naive value, by roughly $10\%$,
$6\%$, and $4\%$ for the three lattice spaces used.  
At two loops there is a further shift,
estimated to be of order (and probably less than) the square of the
first order shift \cite{Oapaper}; around $1\%$, $.36\%$, and $.2\%$ 
respectively.  The latter two are
negligible compared to the statistical errors,
even after taking account of the 5'th power dependence.  It is less
clear how to estimate the importance of $O(a^2)$ nonrenormalizable
operators; but if I estimate all $O(a^2)$ errors by using the three lattice
spacings to extrapolate to $a = 0$, assuming errors proportional
to $a^2$, the result lies within the error bars of the two finer lattice
data, and the error in the extrapolation is dominated by the error in
the finer lattice data.  Hence I will adopt the middle lattice spacing
result and its statistical error bars as the best estimate. 

\section{Interpretation and systematic corrections}
\label{interpretation}

The numerical result is that 
\beq
\Gamma = (10.8 \pm 0.7) \left( \log \frac{m_D}{g^2 T} + O(1) \right) 
	\frac{g^2 T^2}{m_D^2} \alpha_W^5 T^4 
	+ ( {\rm higher \; order \; in \; } g) \, ,
\eeq
but it remains to determine or estimate the rate at the realistic
standard model value of $m_D$, $m_D^2 = (11/6) g^2 T^2$ with $g^2 \simeq
0.4$.  For this value, $\log(m_D / g^2 T) \simeq 1.5$, and the $O(1)$
correction may be quite important.

In fact we might expect that the log needs to be quite large before it
dominates the $O(1)$ ``correction''.  The leading log behavior is based
on the hard particles propagating only a short distance before
undergoing a collision which randomizes their charge.  This ``short''
distance is $4 \pi / (N g^2 T \log(m_D/g^2 T) )$, with $N=2$ since we
are in SU(2) theory.  The leading log contribution to the free path is
$\sim (6 / g^2 T) / \log(m_D/g^2 T)$.  For comparison,
above I confirm Ambj{\o}rn and Krasnitz' result \cite{AmbKras}
that a lattice only $8 / g^2 T$ across is
already large enough to give continuum like behavior for the sphaleron
rate.  The nonperturbative length scale characterizing baryon number
violating processes must be shorter than this, perhaps by a factor
of two. Hence the log will need to be quite big before the ``short''
distance really is short compared to the scale which is setting the
physics.  This supports the expectation that there will be large
corrections to the leading log.

\subsection{estimate using Laine and Philipsen's results}

As I noted already in subsection \ref{HTL_Wilson_subsec}, the log arises
from the behavior of a Wilson line, and the same Wilson line appears in the
definition of the Debye mass beyond leading order.  In this context the
value for the $O(1)$ correction to the leading log behavior has been
found by Laine and Philipsen.  In that case, $[ \log(m_D /
g^2 T) + O(1)]$ has $O(1) = 6.7$.  If the same number held for the
sphaleron rate, then using the standard model value of $m_D$ to evaluate
the log, the leading behavior would be $\Gamma = 89 (g^2 T^2 / m_D^2)
\alpha_W^5 T^4 = 48 \alpha_W^5 T^4$.  This is a crude way of estimating
the nonleading corrections, though, and I do not take it too
seriously.  In particular there is evidence that the length scale
relevant for baryon number violating processes is longer than the
$\simeq 1 / g^2 T$ Laine and Philipsen find for the $O(g^2 T)$
correction to Debye mass; the baryon number violation rate on a cubic
toroidal lattice $3 / g^2 T$ across is over 1000 times slower than for a
large volume\cite{broken_nonpert},  so physics must be going on
involving lengths at least half as long as $3 / g^2 T$.

\subsection{estimate using Hu, M\"{u}ller, and Moore's results}

Another way of trying to determine the subleading corrections is to
use the value of the leading log coefficient to correct my results with
Hu and M\"{u}ller.  There, we used a technique which included hard
thermal loops, but in a way which does not correctly reproduce the
subtleties of the Wilson line responsible for the logarithmic dependence
of $\Gamma$ on $m_D$.  As discussed in subsection \ref{HuMuller_subsec}, 
the log arises 
because excitations more UV than the $g^2 T$ scale make the Wilson
line effectively randomize the charge of a propagating particle,
and there is a log in the reciprocal length for randomization.
The best approach would be to compute the angle averaged value of the
reciprocal length, in the quantum theory and in the lattice theory
actually studied.  Unfortunately, so far I can only compute the
reciprocal length in the lattice theory in one direction, and only in
the cases $m_D \ll 1/a$ or $m_D \gg 1/a$.  The data in \cite{particles}
are taken at $m_D \sim 1/a$, not $\gg 1/a$, so it is at least reasonable
do the match  using the small $m_D a$ approximation.  I will make do
with the log evaluated in the one direction where I can do the integral,
which gives a 
difference in logs between the two theories of $\log(m_D({\rm lattice})
/ m_D({\rm continuum}) + 0.59$.  

I will use the datapoint from that paper taken with the largest value of
$m_D$, because the ``wrong'' lattice mode induced HTL's really are
strongly subdominant to the ``right'' particle induced HTL's for this
case.  Redoing the match between the
lattice and continuum time scales, which was performed wrongly there
because we did not have the $O(a)$ calculation performed in Appendix
\ref{AppendixA} of this paper, revises the result from $\Gamma = 53
\pm 5 (g^2 T^2 / m_D^2) \alpha_W^5 T^4$ down to $50 \pm 5 (g^2 T^2 /
m_D^2) \alpha_W^5 T^4$.  This result was obtained at $m_D 
\simeq 4 g^2 T$, so the difference of logs between the lattice theory
where this number was computed and the quantum theory at the physical
value of $m_D^2$ is about
$.63$ due to the values of $m_D$ plus $.59$ due to UV contributions
present on the lattice but not in the continuum theory.  
Using the determined coefficient of the leading log term, I
should correct the diffusion constant we found downwards by $1.22 \times
10.8$, giving $37 \pm 5$, with only the statistical error bar shown.  
(Substituting in $m_D^2 = 11 g^2 T^2 / 6$ gives
$\Gamma = 20 \pm 3 \alpha_W^5 T^4$.)  

Using $\kappa'$ to correct the old data in this way assumes that 
\begin{equation}
\frac{m_D d}{dm_D} \left( \frac{m_D^2}{g^2 T^2} \Gamma \right) 
	= \kappa' \alpha_W^5 T^4 \, .
\end{equation}
We really only know that this true in the large $(m_D / g^2 T)$ limit,
where the leading log expansion is valid.  It may have quite
nonnegligible corrections at realistic values of $m_D$, which would
appear in a systematic expansion in $\log(m_D/ g^2T)$ as inverse powers
of the log; the $O(1)$ correction to the leading log behavior would
really be $O(1) = C_1 + C_2 (\log(m_D/g^2T))^{-1} + \ldots$.
Intuitively I expect that the real $m_D$ dependence will be weaker than
the leading log suggests (meaning $C_2 > 0$), though I cannot give a
cogent argument to show  this is so.  In this case I have performed an
overcorrection, and the real rate would be higher.  For now I will
accept the corrected answer as the best current guess, but I take a
systematic error bar of order $30\%$ to cover both the rotational
non-invariance not handled correctly in the correction, and the fact
that the correction may be an overestimate at realistic $m_D$.

\subsection{corrections which are formally parametrically suppressed}
\label{sub_in_g}

There is a further cause of systematic error in the determination of the
sphaleron rate, arising from corrections to the $g \ll 1$
approximation.  I will mention the two such corrections which I think
are the largest; fortunately they have opposite sign and the optimistic
can hope that they largely cancel.

One problem is that the parametric argument for the importance of hard
thermal loops assumes $m_D \gg g^2 T$, and it is not clear realistically
whether this is obtained \cite{KLRS_mD}.  There is evidence that it is
not.  In particular, while the sphaleron rate in classical Yang-Mills
theory depends on the lattice spacing in a way consistent with the
Arnold-Son-Yaffe prediction $\Gamma \propto m_D^{-2} \propto a$,
the corrections to the linear dependence are large\cite{slavepaper}.  
Of course, some dependence is expected, since we know now that
the scaling behavior should be not $\Gamma \propto a$ but $\Gamma
\propto a ( \log(1/a) + O(1))$; but we can determine the coefficient of
the $\log(1/a)$ term by using the results of this paper.  According to
Arnold \cite{Arnoldlatt}, we can relate the classical lattice results to
continuum ones, approximately, by replacing $m_D^2 \simeq .684 g^2
T/a$.  In this case the $a$ dependence of $\Gamma$ in classical, lattice
Yang-Mills theory should be
\beq
\Gamma_{\rm on \; the \; lattice} = .465 \kappa' \left( \log \left[ 
	(g^2 aT)^{-1/2} \right] + O(1) \right)  
	\frac{g^2 aT}{4} \alpha_W^4 T^4 \, ,
\eeq
with $\kappa'$ the same as the one we find but the $O(1)$ correction
different.  This formula makes it possible to correct the data in
\cite{slavepaper} to remove the logarithmic dependence on $a$, for
instance by adjusting the data so they all correspond to $g^2 aT =
1/4$, which is the value for the finest lattice used there.  I have done
so, and the result is plotted in Figure \ref{new_fig}.

\begin{figure}
\centerline{\epsfxsize=5in\epsfbox{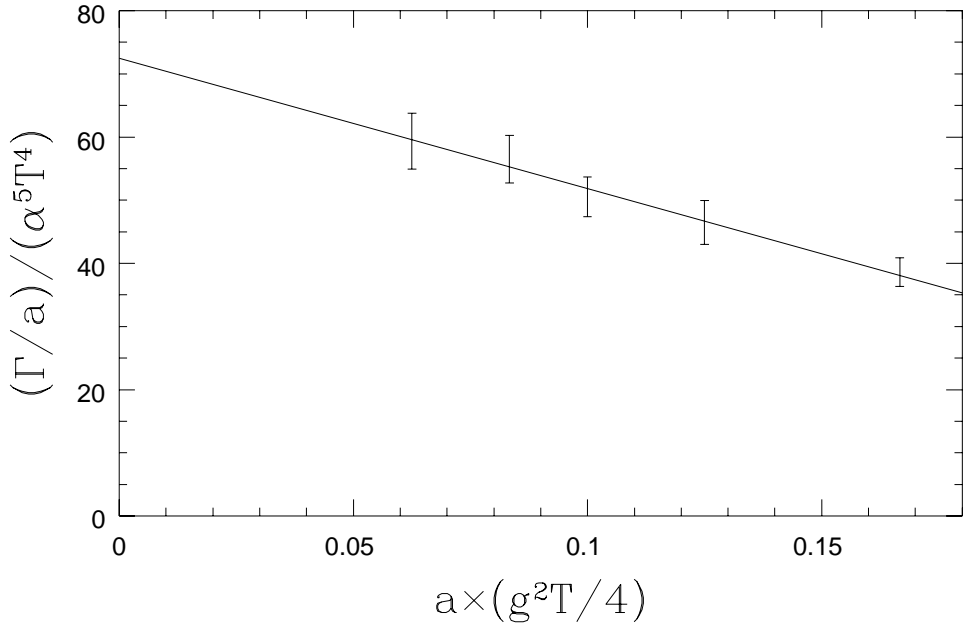}}
\caption{\label{new_fig}  Small lattice spacing extrapolation of data in
pure classical lattice Yang-Mills theory, taken from
{\protect{\cite{slavepaper}}}.  The data have been corrected to absorb the
leading log dependence on $m_D^2 \propto 1/a$ determined here.  They
show a substantial linear correction to the predicted $\Gamma
\propto a$ behavior.  This is evidence of nonnegligible corrections to the
parametric $m_D \gg g^2 T$ limit.}
\end{figure}

I fit the data to the form
$\Gamma / a = C_1 + C_2 a$ to find the corrections to the small spacing
limit which do not arise from the log.
The fit is very good, see Figure \ref{new_fig}, but the coefficient
$C_2$ is quite substantial.  The physical value of HTL strength, using
$m_D = (11/6) g^2 T^2$ and $g^2 = 0.4$, corresponds to $g^2 aT/4 =
.037$, where the correction $C_2 a$ accounts for an $11\%$ shift from
the small $a$ limit, with the actual value falling below the large $m_D$
extrapolation.  The correction reduces the sphaleron rate.
Given the other systematics in play it is probably not fair to call this
a measurement of the correction to the $m_D \gg g^2 T$ limit; rather I
will call it an estimate to tell how large the systematic error is.

Another correction which is parametrically suppressed but not
necessarily very small arises from QCD scatterings of quarks.  Quarks
are responsible for almost 
half of $m_D^2$, and hence almost half of all HTL effects.  But quarks
scatter strongly.  While a strong scattering does not disturb a quark's
electroweak charge, it does change its direction of flight, whereas the
calculation of the HTL's is made assuming particles maintain straight
line trajectories.  If the free path for strong scattering were $\leq 1 /
g^2 T$ this would make an $O(1)$ correction to the quarks' influence on
IR physics.  Actually the quark free path for large angle scattering 
is parametrically order $
( \alpha_s^2 \log(1/g_s) T )^{-1}$ and the correction is formally
$O(\alpha_s^2 / \alpha_W) \sim g^2$; but numerically this might not be
small.  The effect of this correction is to increase the sphaleron rate.
It might be possible to include this correction within the context of
the ``particles'' approach by calculating more precisely the size and
frequency of quark scatterings and adding them to the dynamics of the
particle degrees of freedom.

\subsection{including the Higgs}

I should also mention that the evaluation of the leading log coefficient
made here was within pure SU(2) Yang-Mills theory, without a Higgs
boson.  In the context of baryogenesis we actually need to know the
sphaleron rate in the presence of at least one Higgs boson, in the
symmetric phase and roughly at the equilibrium temperature.  The Higgs
field's evolution is not overdamped, because the hard thermal loops for
a scalar field are nothing but a mass squared correction.  Hence the
$k \sim g^2 T$ modes of the Higgs fields evolve on the time scale $1 /
g^2 T$, which is parametrically faster than the gauge fields.
Therefore, on the $1 / g^4 T$ time scale, the gauge fields only see the
thermodynamic average over all Higgs fields in the fixed gauge field
background\footnote{Dam Son pointed this out to me.}.  
This could be simulated by including the Higgs field in the
Hamiltonian, and evolving it with heat bath dynamics, but updating the
Higgs fields much more often than the gauge fields.  Then we would have
to extract the limit as the Higgs field update is made infinitely faster
than the gauge field update.  I have not yet attempted to do this, but I
anticipate that for parameters which make the phase transition strongly
first order, the effect should be a slight reduction of the leading log
coefficient for the sphaleron rate.

\section{Conclusions}

In the formal small $g$ limit, in which an expansion in $\log(1/g) \gg
1$ is justified, the sphaleron rate in SU(2) Yang-Mills theory is
\beq
\Gamma = (10.8 \pm 0.7) \left( \log \frac{m_D}{g^2 T} + O(1) \right) 
	\frac{g^2 T^2}{m_D^2} \alpha_W^5 T^4 \, .
\eeq
The value $10.8 \pm 0.7$ is clean; the errors are dominated by
statistics, systematics are well under control.

However, interpreting this result to get the sphaleron rate at the
realistic values of $m_D^2 = (11/6) g^2 T^2$ and $g^2 \simeq 0.4$ is
very problematic, because the $O(1)$ correction is not subdominant.  
The reason is that the expansion in large $\log(1/g)$ corresponds to
treating the length scale $2 \pi / [ g^2 T \log(m_D/g^2 T)]$ as much
shorter than the scale relevant for baryon number violating processes,
which is around $\sim 4 / g^2 T$.  The best current estimate for the
$O(1)$ correction comes from using the result of this paper to correct 
previous results of Hu, M\"{u}ller, and myself, which effectively were
using the wrong value for the log.  The result at the physical values for
$m_D$ and $g^2$ is $\Gamma \simeq (20 \pm 3) \alpha_W^5 T^4$, with only
statistical errors quoted.  Unfortunately the correction procedure is
not well under control; I estimate that the systematic errors should be
taken to be at least twice as large as the statistical ones.  There 
are also corrections to the $g \ll 1$ approximation, probably at around
the $10 \%$ level.

Besides making it possible to correct previous results, the leading log
approximation and B\"{o}deker's effective theory provide a 
clean test-bed for determining how good an approximation it is to treat
the symmetric phase of Yang-Mills Higgs theory using just 
Yang-Mills theory.  Checking how much difference the Higgs field
makes is an interesting project for the future.  It would also be
interesting to study the dependence of $\kappa'$ on the number of
colors, since the sphaleron rate in the strong sector is also
phenomenologically interesting for baryogenesis.

\medskip

\centerline{\bf Acknowledgments}

\medskip

I am grateful to Peter Arnold, Dietrich B\"{o}deker, Dam Son, and Larry
Yaffe for illuminating discussions.

\appendix
\section{$O(a)$ lattice time scale renormalization}
\label{AppendixA}

In this appendix I compute at $O(a)$ the lattice to continuum
correction to the Langevin time scale, and the relation between the
continuum Langevin time scale and the amount of heat bath applied.  
The calculation of the Langevin time scales will be quite
technical and will depend to some extent on previous work relating the
lattice and continuum theories at the thermodynamic level, 
found in \cite{Oapaper}.  I will relate the
heat bath depth to the continuum Langevin time scale by making a
nonperturbative lattice measurement comparing it to the lattice Langevin
time scale, and then using the analytic relation between this and the
continuum Langevin time scale.  It would also be possible in principle
to make the connection directly, but the analysis is difficult and I am
lazy. 

\subsection{lattice and continuum Langevin time scales}

Reference \cite{Oapaper} explains how the lattice to continuum relation for
thermodynamical properties of 3-D gauge or gauge-Higgs theory
can be studied as an expansion in $g^2 a T$,
where $a$ is the lattice spacing and $g^2$ is the coupling in 4-D
notation.  The combination $g^2 T$ is the coupling constant of the 3-D
theory.  Since it is dimensionful, a perturbative matching between
continuum and lattice theories must be an expansion in $g^2 T a$ on
dimensional grounds.  In Yang-Mills Higgs theory with fundamental or
adjoint scalar fields the leading terms can behave as $1/a$ because the
theory contains dimension 2 operators, but in strict Yang-Mills theory
the lowest order operator is dimension 4 and the leading corrections are
$O(a)$.  Further, the only $O(a)$ correction is a rescaling of the
coupling, equivalent to a rescaling of the physical length scale, and it
only arises from one loop diagrams, and has been computed
\cite{Oapaper}.

Arnold, Son, and Yaffe have recently demonstrated that, because of
super-re\-nor\-mal\-iz\-a\-bil\-ity and general considerations worked out by
Zinn-Justin and Zwanziger \cite{ZinnJustin}, the same applies to the
dynamics under Langevin dynamics~\cite{ASY2}.  In
particular, any high dimension corrections which can appear in
Eq. (\ref{Effect_theory}) would change the dynamics at $O(a^2)$, simply
because there are no gauge invariant, P even 
dimension 5 operators in Yang-Mills theory.  The
only possible $O(a)$ correction, besides the thermodynamic one already
mentioned, is a rescaling of the Langevin time scale, which will arise
exclusively at one loop.  This opens
the possibility of computing the $O(g^2 aT)$ corrections between the
continuum and lattice Langevin dynamics, by performing a one loop
match.  It also means the Langevin dynamics can be replaced by heat bath
dynamics at the cost of another $O(g^2 aT)$ shift, which I will measure,
rather than compute, in the next subsection.

In the continuum theory and in temporal gauge the Langevin evolution is
\beq
\frac{d A_i^a(x)}{d\tau} = - \frac{\partial H}{\partial A^a_i(x)}
	+ \xi^a_i(x) \, , \qquad \langle \xi^a_i(x,\tau)
	\xi^b_j(y,\tau') \rangle 
	= 2 T \delta_{ij} \delta_{ab} \delta^3(x-y) \delta(\tau - \tau')
	\, . 
\eeq
The Langevin equation for the lattice theory must be written in terms of
the unitary parallel transporter matrices $U_i(x)$ and their derivatives.
By definition the matrix $U_i(x) \in {\rm SU}(N)$ is the matrix such
that, if $\Phi$ is a fundamental representation object transforming as
an object at point $x + a \hat{i}$, $U_i(x) \Phi$ is the parallel
transport to point $x$; thus $U_i(x)$ ``lives on'' the link between the
site $x$ and the site $x + a \hat{i}$.
Writing $D^a_L$ for the left acting derivative,
\beq
D^a_L U = iga T^a U \, , \qquad
D^a_L F(U) = F( U \rightarrow U + D^a_L U ) - F(U) \, ,
\eeq
the Langevin equation for $U$ is
\begin{equation}
\frac{d U_i(x)}{d \tau} = - D^a_L U_i(x) ( \beta_L T D^a_L H_{\rm KS} 
	+ \xi_i^a(x) ) \, , \qquad 
	\langle \xi^a_i(x) \xi^b_j(y) \rangle 
	= \frac{8 \delta_{ij} \delta_{ab} \delta_{xy}}{g^2 a^4 \beta_L} 
	\delta(\tau - \tau') \, .
\label{Latt_langevin}
\end{equation}
Here $T^a$ is a fundamental representation Lie algebra generator with
the standard normalization;
$H_{\rm KS}$ is the Kogut Susskind Hamiltonian, the sum over
elementary plaquettes of the trace of the plaquette,
\beq
H_{\rm KS} = \sum_{\Box} 1 - \frac{1}{2} {\rm Tr} U_{\Box} \, ;
\eeq
and $\beta_L$ is the inverse temperature in lattice units.  At leading
perturbative order $\beta_L = 4
/ ( g^2 a T)$, but it receives radiative corrections, computed in
\cite{Oapaper}, which shift it by a constant.  What I write here as
$\beta_L$ is $\beta_{\rm naive}$ in the notation of \cite{Oapaper}, but
in the body of the paper I have always used the $O(a)$ improved
definitions when I refer to $a$ or $\beta_L$.
The combination $\beta_L H_{\rm KS}$ equals $H/T$ of the continuum
theory, up to radiative corrections and high dimension operators which
correct infrared physics at $O(a^2)$.  The radiative corrections are
absorbed up to errors which are $O(a^2)$ by the one loop radiative
correction to $\beta_L$, which I will present in due course.

Our task is to compute to $O(a)$ the relation between the
Langevin time scale $\tau$ in the continuum and lattice cases.  The
correction at loop order $l$ will be $O(g^2 a T)^l$, 
with the $g^2 T$ from loop counting and the $a$ to balance dimensions;
so we need only go to one loop order.  Part of the 
correction is from the shift in $\beta$ already mentioned, but there are
also corrections from the relation between $U$ and $A$ and from
radiative differences between the lattice $D^a$ and the continuum
derivative.  I will be satisfied to perform the
calculation in a particular gauge, strict Coulomb gauge, without
checking for the gauge independence of the result.  By strict Coulomb
gauge I mean that at every Langevin time the 3-D configuration is in
3-D Landau gauge.  I should also fix a global time dependent gauge
ambiguity; but this is irrelevant at the level of perturbation theory.

In Coulomb gauge the $U$ matrices are all close to the identity matrix,
with a departure of order $\beta_L^{-1/2}$; so it makes sense to define
a lattice gauge field $A_i^a$ through
\beq
U_i(x) \equiv \exp( i g a T^a A^a_i ) \, .
\eeq
the $A$ field so defined has the same normalization as the continuum $A$
field at tree level, but there are radiative corrections.  The value of
the radiative correction to $A$ is obtained in the theory with an added
scalar field by matching the one loop values of the gauge-scalar three
point vertex at small transfer momentum.  There is a contribution from
one loop vertex corrections and from one loop scalar wave function
corrections, and neither depends on the number of scalar particles in
the theory, so the answer is the same in the pure gauge theory.
The one loop correction
was computed as a byproduct in \cite{Oapaper} in a general gauge, for
SU(2). The result, in Coulomb gauge but for SU($N$) gauge theory, is
\beq
A_i^a({\rm continuum}) = \left[ 1 + \left( \frac{Ng^2 a T}{4} \right) 
	\left( \frac{1}{18} \, \frac{\Sigma}{4 \pi} + 3 
	\frac{\xi}{4 \pi} \right) \right] A_i^a({\rm lattice}) \, .
\eeq
The numerical constants $\Sigma$ and $\xi$ were first defined in
\cite{FKRS}, and their numerical values are $\Sigma = 3.175911536$ and 
$\xi = 0.152859325$.

Next I need to find the relation between applying the lattice and
continuum derivatives.  For this purpose it is actually more convenient
to write the update rule in terms of center acting derivatives,
\beq
D^a_C U = U^{1/2} (i g a T^a) U^{1/2} \, ,
\eeq
because the formulation will then be parity symmetric.  Naively one would
expect that if we set $dU_i/d\tau = E^a_i D^a_C U$, that 
$dA_i^a/d\tau = E^a_i$,
and indeed this is correct at leading order in $A$.  But beyond leading
order there are corrections; expanding both sides gives
\beqa
\frac{dU_i}{d\tau} & = & \frac{d}{d\tau} \exp(i g a T^a A^a_i) = 
	iga T^a \frac{dA_i^a}{d\tau}
	- \frac{g^2 a^2 T^b T^c}{2} \left( A_i^b
	\frac{dA_i^c}{d\tau} + (b \leftrightarrow c) \right) \nonumber \\
& &	- \frac{i g^3 a^3 T^d T^e T^f}{6} \left( 
	A_i^d A_i^e \frac{dA_i^f}{d \tau} + {\rm permutations} \right) 
	+ O(a^4) \, , \\
E^a_i D^a U & = & \left[ 1 + iga T^a A^a_i - \frac{g^2 a^2}{2}
	T^b T^c A^b_i A^c_i + O(a^3) \right] i g a T^d E^d_i 
	\times \nonumber \\ & &  \left[
	 1 + iga T^e A^e_i - \frac{g^2 a^2}{2}
	T^f T^h A^f_i A^h_i + O(a^3) \right] \, .
\eeqa
We must expand to $O(a^2)$ corrections because $\langle A^2 \rangle \sim
1/a$.  Equating the expressions, after some work we obtain
\beq
\frac{dA_i^a}{d\tau} ({\rm lattice}) = E_i^d \left[ \delta_{ad} - 
	\frac{g^2 a^2}{24} f_{abe} f_{cde} A_i^b A_i^c 
	+ O(a^3) \right]\, .
\label{A_to_E}
\eeq
To save some writing I left off marking that all $A,E$ above
are in position space and at the same coordinate position.  The above
correction is strictly a lattice effect and the equivalent continuum
relation is $dA_i^a/d\tau = E_i^a$.  

Naively I should now equate $E_i^a$ with 
$- \beta_L D^a_C H_{\rm KS} + \xi$, evaluate the (UV dominated) mean value
of the correction between the lattice and continuum relations, and
thereby determine the rescaling of the Langevin time.  This would not be
strictly correct, though, since the Langevin equation,
Eq. (\ref{Latt_langevin}), gives time evolution in temporal gauge.
The temporal gauge evolution breaks the Coulomb gauge condition, and if
we are to do the calculation in Coulomb gauge we must make a time
dependent gauge change to maintain the Coulomb condition at all Langevin
times.  There is some danger that the gauge changing will also lead to a
radiative correction to the time relation; it turns out it does not, but
I will go through the calculation anyway to show that it does not, since
the cancellations may be special to strict Coulomb gauge.

The Coulomb gauge condition is $D_L \cdot A^a (x) = 0$,
which means
\beq
\sum_i A^a_i(x) - A^a_i(x-a \hat{i}) = 0 \, .
\label{Coul_condition}
\eeq
This does not look cubic invariant because our labeling associates
$U_i$, and hence $A_i$, on the link between $x$ and $x+a \hat{i}$ with
the site $x$; it might better be associated with the point $x + a
\hat{i} / 2$, which would make the cubic invariance more obvious.

To maintain this Coulomb condition, 
we must apply a time dependent gauge transformation at
each site.  The difference between Coulomb gauge and temporal gauge 
satisfying the Coulomb gauge condition at 
$\tau = 0$ will be a gauge transformation by 
$\Lambda = {\cal T} \exp \int ig T^a G^a(\tau) d\tau)$, where the ${\cal T}$
means the exponential should be time ordered 
with respect to $\tau$.  The value of $G$ is fixed by the requirement
that the Coulomb condition remain true; we must choose $G$ so the
departure from the gauge condition due to evolution of the fields and
due to the action of G cancel.  In an infinitesimal time interval
$d\tau$ the gauge change alters $U$ through
\beq
\delta U_i(x) = ( 1 - i g T^a G^a(x) \delta \tau ) U_i(x) ( 1 +
	i g T^b G^b(x+a \hat{i}) \delta \tau ) - U_i(x) \, .
\eeq
Expanding $U$ on each side, the $G$ contribution to the time evolution
of $A$ is
\beqa
\frac{d A_i^a}{d \tau}({\rm from \;}G) & = & \frac{G^d(x+a \hat{i})
	- G^d(x) }{a} \left( \delta_{ad} + \frac{g^2 a^2}{12}
	f_{abe} f_{cde} A^b_i(x) A^c_i(x) \right) - \nonumber \\ & & -
	g f_{abc} A^b_i(x) \frac{G^c(x+a \hat{i}) + G^c(x )}{2} \, .
\label{awful1}
\eeqa

Now we need to substitute this expression, and the relation between $E$
and $dA/d\tau$, into Eq. (\ref{Coul_condition}) to determine the
relation between $E$ and $G$.  The result, Fourier transformed to
momentum space, is
\beqa
0 & = & i \tilde{k}_i E^a_i(k) - \frac{i g^2 a^2}{24} \tilde{k}_i
	f_{abe}f_{cde} \int_{lm} A_i^b(l) A_i^c(m) E_i^d(k-l-m) - 
	\nonumber \\ &&
	- \tilde{k}^2 G(k) - i g f_{abc} \tilde{k}_i \int_l
	\cos(a l_i/2) A_i^b(k-l) G^c(l) - \nonumber \\ &&
	- \frac{g^2 a^2}{12} \tilde{k}_i \tilde{k}_i f_{abe}
	f_{cde} \int_{lm} A_i^b(l) A_i^c(k-l-m) G^d(m) \, .
\label{awful2}
\eeqa
Here summation over vector indices is implied in terms where the index
appears at least 2 times.  I have used the conventional shorthand
$\tilde{k}_i = (2/a) \sin(k_i a/2)$, $\tilde{k}^2 = \sum_i \tilde{k}^2_i$,
and $\int_l = \int d^3 l/(2 \pi)^3$,
where the range of the integration is $[-\pi/a , \pi/a]^3$.
Indices are summed whenever the index appears an even number of times,
which can be more than twice in cubic symmetric but rotationally
nonsymmetric expressions; but if an index appears an even number of
times on each side of a $+$ or $-$ sign I am re-using it, it is an
independent index in each expression.  
To get the continuum version of this expression, drop the $O(a^2)$ terms
and set $\tilde{k} = k$, $\cos(ka/2) = 1$.

We can determine $G$ perturbatively in $g$ by expanding
Eq. (\ref{awful2}) in powers of $g$.  The result to $O(g^2)$ is
\beqa
G^a(k) & = & i \frac{\tilde{k}_i E_i^a}{\tilde{k}^2}
	+ g f_{abc} \int_l \frac{\tilde{k}_i \tilde{l}_j}{\tilde{k}^2 
	\tilde{l}^2} \cos(\tilde{l}_i a/2) A_i^b(k-l) E_j^c(l) -
	\nonumber \\ & & 
	- i g^2 f_{abe}f_{cde} \int_{lm} \frac{\tilde{k}_i \tilde{l}_j
	\tilde{m}_k}{\tilde{k}^2 \tilde{l}^2 \tilde{m}^2} 
	\cos(\tilde{l}_ia/2) \cos(\tilde{m}_ja/2) A_i^b(k-l)
	A_j^c(l-m) E_k^d(m) - 
	\nonumber \\ & & 
	- \frac{i g^2 a^2 \tilde{k}_i}{24 \tilde{k}^2} f_{abe} f_{cde}
	\int_{lm} A_i^b(l) A_i^c(m) E_i^d(k-l-m) -
	\nonumber \\ & & 
	- \frac{ig^2 a^2}{12} f_{abe} f_{cde} \int_{lm}
	\frac{\tilde{k}_i \tilde{k}_i \tilde{m}_j}
	{\tilde{k}^2 \tilde{m}^2} A_i^b(l) A_i^c(k-l-m) E^d_j(m) \, .
\label{awful3}
\eeqa
This in turn must be substituted into Eq. (\ref{awful1})
to find the true value of $dA/d\tau$ in Coulomb gauge,
\beqa
\frac{dA_i^a(k)}{d\tau} & = & \left( \delta_{ij} - \frac{\tilde{k}_i 
	\tilde{k}_j}{\tilde{k}^2} \right) E^a_j 
	+ i g f_{abc} \int_l \frac{\tilde{l}_k}{\tilde{l}^2} 
	\left(\delta_{ij} - \frac{\tilde{k}_i \tilde{k}_j} 
	{\tilde{k}^2}\cos(l_ja/2) \right) A_j^b(k-l) E_k^c(l) +
	\nonumber \\ & & 
	+g^2 a^2 f_{abe} f_{cde} \int_{lm} \Bigg[ \frac{-1}{24}
	\left( \delta_{ij} - \frac{\tilde{k}_i \tilde{k}_j}
	{\tilde{k}^2} \right) A_j^b(l) A_j^c(m) E_j^d(k-l-m) -
	\nonumber \\ & & \qquad - \frac{1}{12} \left( \frac{\tilde{m}_i 
	\tilde{m}_k \delta_{ij}}{\tilde{m}^2} - 
	\frac{\tilde{k}_i \tilde{k}_j \tilde{k}_j \tilde{m}_k}
	{\tilde{k}^2 \tilde{m}^2} \right) A_j^b(k-l-m) A_j^c(l) 
	E_k^d(m) - \nonumber \\ & & \qquad
	- \frac{\tilde{l}_k \tilde{m}_l}{\tilde{l}^2 \tilde{m}^2 } 
	\left( \delta_{ij} - \frac{\tilde{k}_i \tilde{k}_j}{\tilde{k}^2} 
	\right) \cos(l_ja/2) \cos(m_ka/2) \times
	\nonumber \\ & & \qquad \qquad A_j^b(k-l) A_k^c(l-m) E_l^d(m) 
	\Bigg] \, .
\label{awful4}
\eeqa

Below I will be interested in the case where $E$ is uncorrelated with
$A$, in which case the mean value of the term linear in $A$ vanishes and
we can substitute the leading order Landau gauge $A$ field correlator, 
\beq
\langle A_i^a(k) A_j^b(l) \rangle = \delta_{ab} \delta(k+l) 
	\left( \delta_{ij} - \frac{\tilde{k}_i \tilde{k}_j}
	{\tilde{k}^2} \right) ( \tilde{k}^2 )^{-1} \, ,
\eeq
into the remaining terms.  The contributions from the terms with (1/12)
and (1) in front vanish, only the term with the leading (1/24)
coefficient contributes.  It requires that we perform the integral
\begin{equation}
a^2 \int_l \frac{1}{\tilde{l}^2} = a \frac{\Sigma}{4 \pi} \, ,
\end{equation}
which is the definition of the constant $\Sigma$ which appeared
earlier.  We also need the integral
\beq
a^2 \int_l \frac{\tilde{l}_1^2}{(\tilde{l}^2)^2} = \frac{a}{3}
	\frac{\Sigma}{4 \pi} \, .
\eeq
The result is
\begin{equation}
\frac{dA_i^a(k)}{d\tau} ({\rm lattice}) = 
	\left( \delta_{ij} - \frac{\tilde{k}_i 
	\tilde{k}_j}{\tilde{k}^2} \right) 
	\left[ 1 + \frac{1}{9} \, \frac{Ng^2 a T}{4} \, 
	\frac{\Sigma}{4 \pi} \right] E_j^a(k) \, ,
\end{equation}
valid when $E$ is uncorrelated with gauge fields.  The first factor
projects out the transverse component of $E$ and is responsible for 
maintaining Coulomb
gauge.  The continuum expression is the same but with $a$ set to zero.

Using the previously
established relation between lattice and continuum gauge field
normalization, the relation for the continuum normalized gauge field is
\begin{equation}
\frac{dA^a_a(k)({\rm contin})}{d\tau({\rm latt})} = 
	\left[ 1 + \frac{Ng^2 aT}{4} \left( \frac{1}{6} \, 
	\frac{\Sigma}{4 \pi} + 3 \frac{\xi}{4 \pi} \right) \right] 
	\left( \delta_{ij} - \frac{\tilde{k}_i 
	\tilde{k}_j}{\tilde{k}^2} \right) E_j^a(k) \, ,
\label{the_result}
\end{equation}
valid for the gauge field response to the random force part of the
Langevin equation.  I will not attempt to study the response of $A$ from
the Hamiltonian gradient part of the Langevin equation, since it would
involve understanding the radiative corrections to $D^a H$ and would
lead to an $E$ which might be correlated with $A$ fields.  Since we know
that the Langevin equation correctly thermalizes the lattice system when
we use the radiatively corrected value of $\beta_L$, it is sufficient to
study the response to the random force alone to determine the rescaling
of the Langevin time scale.  

Note that the correction appearing in
Eq. (\ref{the_result}) is precisely one quarter of the radiative 
wave function correction for an adjoint scalar field in 3-D lattice gauge
theory when the scalar self-coupling vanishes, see \cite{Oapaper}.  
This is perhaps not too surprising.  In the real time theory, the time
evolution is generated by the electric fields, which appear in the
thermodynamics as the $A_0$ field, an adjoint scalar with zero
self-coupling.  In \cite{slavepaper} Turok and I speculated incorrectly
that the time scale correction would not contain any large tadpole
corrections; but Eq. (\ref{the_result}) is one quarter the adjoint scalar
wave function correction, which does contain tadpoles.  Our incorrect
guess was based on analyzing the abelian theory, 
where the $A_0$ field turns out to be a free field.  
The correction found here vanishes in the abelian case, for compact or
noncompact implementations.

Now I will finish the relation between time scales.  I have just shown
that the response of the $A$ field to the random force, 
shifting its normalization to correspond to the continuum theory
normalization, is
\beq 
\frac{dA_i^a(x)}{d\tau({\rm latt})} = ( 1 + {\rm corr} ) \xi \, ,
\eeq
where $(1+{\rm corr})$ is the quantity in brackets in
Eq. (\ref{the_result}).  But the autocorrelator of $\xi$ is not the same
as it would be in the continuum, because of the radiative corrections to
$\beta_L$.  The mean square change in $A_i^a$ over a Langevin time
$\delta \tau({\rm latt})$ is, using Eq. (\ref{Latt_langevin}),
\beq
\langle (\delta A_i^a(x))^2 \rangle = (1 + {\rm corr})^2 
	\frac{2 T \delta \tau({\rm latt}) }{a^3}
	\frac{4}{g^2 a T \beta_L} \, . 
\eeq
However, $\beta_L = 4 / g^2 a T$ only at leading order in $\beta_L$.
Beyond leading order, in the pure gauge theory, it is \cite{Oapaper}
\begin{equation}
\frac{\beta_L g^2 a T}{4} - 1 \equiv Z_g^{-1} - 1 = \frac{g^2 aT}{4} \left( 
	\frac{N^2 - 2}{3N} + \frac{37N}{3} \, \frac{\xi}{4 \pi}
	\right) \, .
\end{equation}
The continuum theory Langevin equation would cause a mean square change
to $A$ of 
\begin{equation}
\langle (\delta A_i^a(x))^2 \rangle = \frac{2 T \delta \tau({\rm 
	contin})}{a^3} \, ,
\end{equation}
so the relation between time scales is 
\begin{equation}
\frac{\delta \tau({\rm contin})}{\delta \tau({\rm latt})} = 
	(1 + {\rm corr})^2 Z_g = 
	 1 + \frac{g^2 a T}{4} \left( \frac{N}{3} \, 
	\frac{\Sigma}{4 \pi} - \frac{19}{3} \, \frac{\xi}{4 \pi} 
	- \frac{N^2 - 2}{3 N}
	\right) \, .
\end{equation}
The numerical value of this expression for $N=2$ is
$1 - 0.3189 (g^2 a T/4)$.

I have now related the lattice and continuum Langevin time scales at
$O(a)$.  It is also easy to show that, for the Hamiltonian system, the
correction between the lattice and continuum time scales 
is exactly half as large.  However this is less useful in
that case because, while this correction is technically correct for
determining the time falloff of correlators over very short time scales,
the IR dynamics on longer time scales receive HTL corrections which
depend on the lattice spacing as $1/a + O(1)$.  For the technique of Hu
and M\"{u}ller, HTL's are included by adding ``particle'' degrees
of freedom.  In \cite{particles} we work out the correction for time
scales in the limit $m_D^2 \gg g^2 T/a$, in an approximation which
corresponds to setting $(1 + {\rm corr}) = 1$.  The correction found
here changes our result there from being $t_{\rm latt} /
t_{\rm contin} = Z_g^{-2}$ 
to being $Z_{g}^{-2} (1 + {\rm corr})^{-2}$.  I used this
correction in Section \ref{interpretation}.

\subsection{lattice Langevin time and depth of heat bath}

The effect of the heat bath algorithm on the infrared degrees of freedom
is equivalent to Langevin evolution.  I will show this at tree
level, which gives a tree relation between Langevin time and the number
of heat bath updates applied.  The only $O(a)$ correction to this
relation possible is an $O(a)$ shift in time scales between the
algorithms; to find the magnitude of the shift I make a direct numerical
measurement.  I do not attempt a
perturbative calculation of the relation between continuum Langevin time
and amount of heat bath applied, beyond leading order.

First I will show that the effect of the heat bath algorithm on the IR
degrees of freedom is equivalent to the Langevin algorithm, and I find
the relation between the Langevin time scale to the number
of links updated by heat bath, at leading order.  The way the heat bath
algorithm works is as follows: 
\begin{enumerate}
\item 	pick a link on the lattice at random.
\item 	replace the connection $U$ on the link with the one which
	minimizes the Hamiltonian.
\item   multiply this link by a random SU(2) element chosen from a 
	distribution centered on the identity, with a weight function 
	dependent only on the
	arc length from the identity, not the direction.  The weight is
	chosen to correctly reproduce the thermal ensemble on this link
	holding others fixed; it is approximately but not exactly
	Gaussian. 
\end{enumerate}
This is a heat bath update; for a more precise description
see \cite{Pendleton}.  The first part, the quench of the link, serves to
perform the $- \partial H/ \partial A$ part of the Langevin update, and the
multiplication by a random SU(2) element reproduces the noise part of
the Langevin update.

To see the relation between the heat bath and the Langevin update at
leading order it is sufficient to consider the linearized theory, that
is, to expand the Hamiltonian to quadratic order in the gauge field
$A$.  On an $N\times N\times N$ toroidal lattice the relation between
the connections $U$, the gauge field $A_i(x)$, and the Fourier transform
of the gauge field $A(k,s)$ ($s$ a polarization index) is (writing all
Lorentz and group index sums explicitly, there is no implicit summation
convention in what follows) 
\beqa
U_i(x) & = & \exp \left( \sum_a iga T^a A^a_i(x) \right) 
	\simeq 1 + \sum_a iga T^a A^a_i(x) -
	\frac{g^2 a^2}{8} \sum_a A^a_i(x) A^a_i(x) \, , \\
\label{A_transform}
A^a(k,s) & = & N^{-3/2} \sum_{x,i} \epsilon_i(s,k) A^a_i(x)
	\exp\left( -i k \cdot(x + a \hat{i}/2) \right) \, , \\
\label{whatAis}
A_i^a(x) & = & ({\rm longitudinal \; piece}) + 
	N^{-3/2} \sum_{k,s} A^a(k,s) \epsilon_i(s,k)
	\exp\left( i k \cdot(x + a \hat{i}/2) \right) \, ,
\eeqa
where $\epsilon_i(s,k)$ is a transverse polarization vector, satisfying
\beq
\label{orthogonality}
\sum_i \epsilon_i(s,k) \epsilon_i(s',k) = \delta_{s,s'} \, ,
\eeq
and 
\beq
\label{transversality}
\sum_i \epsilon_i(s,k) \tilde{k}_i = 0 \, ; 
\eeq
there are two such states for
each $k$.  The sum over $k$ includes all $k$ of form $(2
\pi/aN)\vec{n}$, with $\vec{n}$ a triple of integers each in the range
$0 \leq n_i < N$.  Only the transverse fields are of interest here, at
the order we are working the longitudinal part is pure gauge.  Its
behavior depends on our choice of gauge fixing, and has no influence on
the transverse fields (or on physics).  (This is not true at higher order
in $g$, where the longitudinal terms are responsible for the Fadeev-Popov
determinant.)
At the level we are working the value in thermal equilibrium of $A$ is
given by 
\beq
\label{equipartition}
\langle (A^a(k,s))^* A^b(k',s') \rangle 
	= \delta_{k,k'} \delta_{ab} \delta_{s,s'} 
	\frac{T}{\tilde{k}^2}
\eeq

Now let us analyze how the fields change when the heat bath algorithm is
applied to a link $(x,i)$.  The
terms in the Hamiltonian containing the link $x,i$ are
\beqa
\frac{H}{T} & \supset & \frac{1}{a^2 T} \sum_{a,j \neq i} \left[
	\frac{1}{2} \left( A^a_j(x) + A^a_i(x+a \hat{j}) - 
	A^a_j(x+a \hat{i}) - A^a_i(x) \right)^2  + \right.
	\nonumber \\ & & \qquad \left.
	\qquad + \frac{1}{2} \left( -A^a_j(x-a\hat{j}) + A^a_i(x-a \hat{j)} 
	+ A^a_j(x-a(\hat{i}+\hat{j})) - A^a_i(x) \right)^2 \right] \, ,
\label{whatHis}
\eeqa
and ``quench'' part of the heat bath algorithm will replace $A^a_i(x)$
with the value which minimizes this expression,
\beqa 
A^a_i(x,{\rm after}) & 
	= & \frac{1}{4} \sum_{j \neq i} \left( A^a_j(x) + A^a_i(x+a \hat{j}) - 
	A^a_j(x+a \hat{i}) - \right. \nonumber \\ & & \qquad \qquad 
	\left. - A^a_j(x-a\hat{j}) 
	+ A^a_i(x-a \hat{j)} + A^a_j(x-a(\hat{i}+\hat{j})) \right) \, .
\eeqa
Using Eq. (\ref{whatAis}) and Eq. (\ref{transversality}), and adding a
term $\xi^a$ to represent the noise which will be added, this is
\beq
\label{A_after}
A^a_i(x) ({\rm after}) = A^a_i(x,{\rm before}) 
	+ \xi^a - \sum_{k,s}
	\frac{a^2 \tilde{k}^2}{N^{3/2}}	\epsilon_i(s,k) A^a(k,s) 
	\exp(i k \cdot (x + \hat{i} a/2)) \, .
\eeq
Now we should compute the size of $\xi^a$.  Because the Hamiltonian is
expanded only to quadratic order, the noise is Gaussian, of amplitude
set by the size of the 
quadratic in $A_i(x)$ term in $H$, which from
Eq. (\ref{whatHis}) is $(2/a^2T)\sum_a A^a_i(x) A^a_i(x)$.  The amplitude of
the noise $\xi^a$ is then (no sum on $a$) 
$\langle \xi^a \xi^a \rangle = a^2T/4$.

Next we will see what impact this update has had on the Fourier mode
$A(k,s)$.  Combining Eqs. (\ref{A_transform}) and (\ref{A_after}), we
find 
\beqa
A^a(k,s,{\rm after}) & \! \! \! = \! \! \! & A^a(k,s,{\rm before}) \left(
	1 - \frac{\epsilon^2_i(s,k) a^2 \tilde{k}^2}{4 N^3} \right)
	+ \frac{\epsilon_i(s,k)}{N^{3/2}} 
	\exp \left( -ik \cdot (x+a\hat{i}/2) \right) \xi^a
	\nonumber \\ & & 
	- \! \! \! \sum_{(s',l) \neq (s,k)} \! \! \frac{a^2 \tilde{l}^2 
	\epsilon_i(s,k) \epsilon_i(s',l)}{4 N^3}
	A^a(l,s',{\rm before}) 
	\exp \left( i(k-l) \cdot (x + a \hat{i}/2) \right) \, .
\label{new_Ak}
\eeqa
The $(k,s)$ term in the sum is removed and included instead in the first
term.  It is responsible for the damping term in the Langevin equation.
Both the noise term $\xi$ and the final term, which I will call the
``cross-talk'' term, are responsible for the noise term in the Langevin
equation.  

To measure the magnitude of the Langevin damping term, we must compute
the correlator  
$\langle A^*(k,s,{\rm before}) \: A(k,s,{\rm after}) \rangle$.  Because 
$\langle A(k,s) \: A(l,s') \rangle = 0 = \langle A(l,s') \: \xi \rangle$, we
get 
\beq
\label{damp_rate}
\langle A^*(k,s,{\rm before}) A(k,s,{\rm after}) \rangle = 
	\langle A(k,s,{\rm before})^2 \rangle \left( 1 - 
	\frac{a^2 \tilde{k}^2 \epsilon^2_i(s,k)}{4 N^3} \right) \, .
\eeq

It is also important to make sure that the mean square value of
$A^a(k,s)$ is unchanged by the update, which is the requirement that the
noise have the right amplitude.  Here we get a little surprise; squaring
Eq. (\ref{new_Ak}), 
\beqa
\langle (A^a(k,s,{\rm after}))^2 \rangle & = & 
	\left( 1 - \frac{a^2 \tilde{k}^2 \epsilon_i^2(s,k)}{4N^3} \right)^2 
	\langle (A^a(k,s,{\rm before})^2 \rangle 
	+ \frac{\epsilon_i^2(s,k)}{N^3} \langle (\xi^a)^2 \rangle
	\nonumber \\
& & + \frac{1}{N^6} \sum_{(l,s') \neq (k,s)} \langle A^a(l,s')
	A^a(l,s') \rangle \epsilon^2_i(s,k) \epsilon^2_i(s',l)
	\left( \frac{a^2 \tilde{l}^2}{4} \right)^2 \, .
\label{horridd}
\eeqa
Using Eq. (\ref{equipartition}) and taking $N^3 \gg 1$, this becomes
\beq
\langle (A^a(k,s,{\rm after}))^2 \rangle =
	\langle (A^a(k,s,{\rm before})^2 \rangle
	+ \frac{a^2 T \epsilon_i^2(s,k)}{N^3} \left( - \frac{1}{2}
	+ \frac{1}{4} + \frac{1}{16 N^3} \sum_{l,s'}
	\epsilon^2_i(s',l) a^2 \tilde{l}^2 \right) \, .
\label{final_noise}
\eeq
At leading order in large $N$, the sum gives $4 N^3$.  Therefore the
mean size of $A^a(k,s)$ is unchanged, which means that we have the
correct amount of noise.  We see
that fully half of the noise actually arises from ``cross-talk'' between
the mode of interest and extremely UV modes, with the other half arising
from the noise explicitly appearing in the algorithm.  

To be Langevin the noise must have zero unequal time correlation.  This
is the case for $\xi$ by explicit construction, but we need to check it
for the ``cross-talk'' noise.  For the heat bath algorithm to act like a
Langevin algorithm, the ``cross-talk'' contribution to $A(k,s)$ from
updating the $(x,i)$ link must be independent of that from the $(y,j)$
link, at least on averaging over the choice of $(y,j)$ (which is indeed
chosen randomly in the algorithm I use).  This is the case; from
Eq. (\ref{new_Ak}), the correlation between the ``cross-talks'' is
\beqa
\! \! & & \sum_{(l,s') \neq (k,s)} \: \sum_{(m,s'')\neq (k,s)}
	\left( \frac{a^4 \tilde{l}^2 \tilde{m}^2}{16 N^6} \right)
	\epsilon_i(s,k) \epsilon_j(s,k) \epsilon_i(s',l)
	\epsilon_j(s'',m) \nonumber \\ & & \hspace{1.4in}
	\times \langle A^a(l,s') A^a(m,s'') \rangle
	\exp \left( i k \cdot (x-y+a(\hat{i}-\hat{j})/2) \right)
	\nonumber \\ & & \hspace{1.4in} \times
	\exp \left( -il \cdot(x+ a \hat{i}/2) 
	+ i m \cdot(y + a\hat{j}/2) \right) \nonumber \\ 
\! \! & \! \! = \! \! & \sum_{(l,s') \neq (k,s)} 
	\frac{a^4 \tilde{l}^2}{16 N^6}
	\epsilon_i(s,k) \epsilon_j(s,k) \epsilon_i(s',l) 
	\epsilon_j(s',l) 
	\exp \left( i (k-l) \cdot (x-y+a(\hat{i}-\hat{j})/2) \right) 
	\, ,
\label{unequal_t_cross-talk}
\eeqa
which suffers from a rapidly oscillating phase.  The expression is
typically smaller in magnitude by $N^{-3/2}$ compared to the
corresponding term in Eq. (\ref{horridd}) and its average over $(y,j)$
is strictly zero.  Hence there is no unequal time correlation in the
``cross-talk'' part of the noise.  Note also that the cross-talk is very
strongly UV dominated, which means that there will be no hidden
correlations in the IR effective evolution because of it, 
at least at $O(a)$ and probably higher; it is also
fortunate because the UV is most quickly randomized.

What we have shown is that applying the heat bath update is equivalent
to damping the $A$ fields and applying noise.  In particular, its
influence on the IR degrees of freedom is equivalent to that of 
Langevin dynamics.  Applying the heat bath algorithm to many links in
succession, the rate at which a mode is damped (and the amount of noise
it receives) is given by Eq. (\ref{damp_rate}) (and
Eq. (\ref{final_noise})) after averaging over the direction $i$ (since
each direction is bathed with equal frequency).  Using $\sum_i
\epsilon_i^2(s,k) = 1$, we find that it 
takes $12 N^3$ heat bath updates to perform the equivalent of $a^2$ of
Langevin update.  This gives the tree relation between the algorithms.
We can then define a heat bath time in
terms of the number of links we have updated, $\tau({\rm heatbath}) = 
a^2({\rm number \; of \; links \; updated})/(12 N^3)$.  
At leading order in
a weak field expansion this is the same as Langevin time but we expect
subleading corrections.

Now we must push the analysis beyond tree level.
Since the influence of the heat bath algorithm on the IR
degrees of freedom (in fact, all degrees of freedom) 
``looks like'' Langevin dynamics, the analysis of \cite{ZinnJustin}
applies; up to high dimension operator corrections, which by power
counting appear first at $O(a^2)$, the algorithms are
related, in the presence of interactions, by a rescaling of all
parameters.  Since each algorithm gives correct thermodynamic behavior
(after the $O(a)$ correction already discussed is applied), the only
remaining correction would be a rescaling of the time scale, which must
be at worst $O(a)$ since it vanishes as $g^2 \rightarrow 0$ (in
which limit the calculation just presented is exact), and on dimensional
grounds any $O(g^2)$ correction must be $O(g^2 aT)$.  

We could in
principle determine this $O(a)$ correction by an analytic computation,
extending the one just presented to second order in $g$.
Instead I compute the subleading effects by the following strategy.
I choose some infrared measurable ${\cal O}$, 
and measure it at each lattice point at a closely spaced series of
Langevin times.  I do the same using the heat bath algorithm.  Then I
compare the unequal time correlator or autocorrelator $C(\tau - \tau') = 
\langle {\cal O}(x,\tau) 
{\cal O}(x, \tau') \rangle$, where the average is over the ensemble of
Langevin trajectories, or in practice over coordinates and times in a
single very long Langevin trajectory.  To match the 
time scales, we see what rescaling of the heat bath time scale
is needed to make the autocorrelator match the autocorrelator for the
Langevin case.  Any IR measurable will do because we know that the
algorithms both behave as Langevin algorithms on the IR degrees of
freedom, so the only $O(a)$ difference would be a time rescaling which
will be of the same amplitude for any unequal time observable.

I should explain that the reason this is worth doing at all is that,
first, the heat bath algorithm is much faster and does not suffer from
step size errors like the Langevin algorithm, and second, there are
infrared measurables other than the topological density, for which the
autocorrelation statistics improve much more quickly.  If the latter
were not true we would spend as much computer time making the match
between techniques as it would take to do the measurement of $\Gamma$ by
the Langevin method.

The measurable I choose is a fundamental representation Wilson loop
after some amount of cooling, specifically a $4 \times 4$ square Wilson
loop after  $\tau = 3.125 a^2$ of gradient flow cooling under $H_{\rm
KS}$.  This is an infrared measurable because such a large Wilson loop
samples mostly the infrared gauge fields, and because the cooling
removes most of the UV fluctuations anyway.  

Incidentally, it is not too hard to compute the
leading order perturbative prediction for this quantity.  The mean trace
of an $l \times
l$ Wilson loop in SU($N$) gauge theory after a length $\tau$ of gradient 
flow cooling, neglecting lattice artifacts, is
\beq
N - {\rm Tr} U_{l \times l} = (N^2-1) \frac{g^2 T}{4}
	\int \frac{d^3 k}{(2 \pi)^3} \frac{ 16 \sin^2 (k_x l/2) 
	\sin^2(k_y l/2)}{k_x^2 k_y^2} \frac{k_x^2 + k_y^2}{k^2}
	\exp( - 2 k^2 \tau) + O(g^4)\, ,
\eeq
In the $\tau \rightarrow 0$ limit the integral has logarithmic UV
divergences but for finite $\tau$ it has a well defined value and is
dominated by the infrared, $k \lsim \tau^{-1/2}$.  Of
course, for the Wilson loops under consideration here, perturbation
theory will be unreliable because the length scales involved are close to
the scale where perturbation theory breaks down completely.  It might be
interesting to see whether the infrared fields are stronger or weaker
than at leading order in perturbation theory, though.

I measured the same site, unequal time Wilson loop correlator by
measuring each $4 \times 4$ Wilson loop in an even sublattice every
$a^2/2$ of Langevin time, for a series of Langevin trajectories each
about $200 a^2$ long, with $50 a^2$ Langevin time between trajectories
to eliminate correlations between trajectories.  For each trajectory I
determined the autocorrelator averaged over volume and time.  
The autocorrelation function $C(\Delta \tau)$ looks something like an
exponential tail but is not well fit by one; there is some small amount
of much longer time scale correlation caused by the slow evolution of the
most infrared gauge fields.  To compare
the Langevin and heat bath time scales I averaged $C(\Delta \tau)$ over
data sets for each update method and determined the rescaling of the
Langevin time which minimized the difference between the results,
\begin{equation}
\chi^2 = \int_0^{\Delta \tau_{\rm max}} ( C_{\rm Langevin}(\Delta \tau) - 
	r_C C_{\rm heat}(r_\tau \Delta \tau) )^2 d \Delta \tau \, ,
\end{equation}
where I allow a rescaling both of the autocorrelation time and of the
overall magnitude of $C$.  I chose $\Delta \tau_{\rm max}$ to be enough
that the autocorrelator has fallen about $1-e^{-1}$ of the way to its
large $\Delta \tau$ limit, but the result is quite insensitive to the
specific choice.  The coefficient $r_\tau$ at the extremum of
$\chi^2$ gives the rescaling of the Langevin time scale.  The
multiplicative rescaling of $C$ is necessary because there are small,
very long time scale correlations in the measurable which can
effectively shift one data set somewhat with respect to the other.
As a check I compared the first half of the Langevin data I took with
the second half.  I find that
$r_\tau=1$ within a small tolerance the same size as the jackknife error
bars, but $r_C$ differs from $1$ by a few percent.  

For $(4 / g^2 aT) = 6$ I ran each update procedure on a $24^3$ lattice
for $\tau \simeq 9000 a^2$.  The rescaling of the time scales was
\beq
\Delta \tau_{\rm heat}  =  1.098 \pm .007 \Delta \tau_{\rm Langevin} 
	\, ,
\eeq
with the error bar determined by the jackknife method.  

The Langevin step size used here was $\Delta = .05$.  The definition of
$\Delta$ and the second order algorithm I used are in Section
\ref{numerics} in the body of the paper.  To check for step size errors
I evolved a trajectory for half as much Langevin time, using $\Delta =
.025$.  The rescaling between this trajectory and the heat bath was
$1.099 \pm .010$ and the rescaling between Langevin evolutions with the
two step sizes was $1.001 \pm .009$, so Langevin 
step size errors are negligible at $\Delta = 0.05$.  

I also measured
$N_{\rm CS}$ during the $\Delta = 0.05$ Langevin trajectory; the ratio
of the diffusion constants for the heat bath and Langevin algorithms was
\beq
\frac{\Gamma_{\rm heat} }{\Gamma_{\rm Langevin}} ({\rm unrescaled})
	= 1.114  \pm .058 \, ,
\eeq
which is compatible with the measured difference in time scales, 
but with much larger error bars.  The autocorrelator
of the Wilson loop develops good statistics more quickly than the
diffusion constant for $N_{\rm CS}$.  

Incidentally the mean value of the
Wilson loop trace was $\langle 2 - {\rm Tr} U_{4 \times 4} \rangle 
= .261$, while the leading order perturbative prediction is $.096$.  The
infrared of Yang-Mills theory has more excitation than leading order
perturbation theory predicts, by quite a bit on the scale of a
square Wilson loop $8/(3 g^2 T)$ on a side.

For $(4/ g^2 a T) = 10$ I used a $40^3$ lattice but only $\tau = 2000
a^2$.  The rescaling of the time scales was smaller as expected, 
$1.065 \pm .016$.  Assuming the subleading correction to be purely
$O(a)$, we would have guessed from the $\beta_L = 6$ result that the
rescaling factor would be $1.059 \pm .004$, which is within error.
The rescaling at both lattice spacings are within error of being
$\tau_{\rm Langevin}/\tau_{\rm heat} = Z_g$, and I speculate that this
is the correct analytic relation at $O(a)$.
For the finest lattice spacing, I have simply extrapolated the medium
spacing data assuming a pure $O(a)$ form for the correction.

The deviation of the mean value of the Wilson loop measurable from
perturbation theory is also smaller on the finer lattice; 
the value is  $.110$, while the perturbative estimate is $.058$.
This is also expected, since a $4 \times 4$ Wilson loop is smaller in
physical units in this case, and perturbation theory works better as the
length scale becomes smaller.

\end{document}